\documentclass[useAMS]{mn2e}
\usepackage{graphicx}
\usepackage{epsfig}
\usepackage{amssymb}
\usepackage{lscape}
\usepackage{ulem}
\usepackage{txfonts}

\def\siggas{$\Sigma_{\rm gas}$}
\def\sigh2{$\Sigma_{\rm H_2}$}
\def\sighi{$\Sigma_{\rm HI}$}
\def\sigsfr{$\Sigma_{\rm SFR}$}

\def\sigstar{$\Sigma_{\star}$}

\def\kms{km~s$^{-1}$}

\def\c2s{C\,{\sc ii}$^{\star}$}

\title[ALMaQUEST XI] {The ALMaQUEST Survey XI: A strong but non-linear relationship between star formation and dynamical equilibrium pressure.}

\author[Ellison et al.] {Sara L. Ellison$^1$,  Hsi-An Pan$^2$, Asa F. L. Bluck$^{3}$, Mark R. Krumholz$^{4,5}$, Lihwai Lin$^6$,   
  \newauthor   Leslie Hunt$^7$, Edvige Corbelli$^7$,  Mallory D. Thorp$^{1,8}$, Jorge Barrera-Ballesteros$^9$, 
  \newauthor     Sebastian F. S\'{a}nchez$^9$, Jillian M. Scudder$^{10}$,   Salvatore Quai$^{11,12}$  \\ 
$^1$ Department of Physics \& Astronomy, University of Victoria, Finnerty Road, Victoria, British Columbia, V8P 1A1, Canada\\
  $^2$ Department of Physics, Tamkang University, No.151, Yingzhuan Road, Tamsui District, New Taipei City 251301, Taiwan\\
  $^3$ Department of Physics, Florida International University, 11200 SW 8th Street, Miami, FL, USA\\
  $^4$ Research School of Astronomy and Astrophysics, Australian National University, Canberra ACT 2600, Australia\\
  $^5$ ARC Center of Excellence for Astronomy in Three Dimensions (ASTRO-3D), Canberra ACT 2600, Australia\\
  $^6$ Institute of Astronomy \& Astrophysics, Academia Sinica, Taipei 10617, Taiwan\\
  $^7$ INAF— Osservatorio Astrofisico di Arcetri, Largo E. Fermi 5, I-50125, Florence, Italy\\
  $^8$ Argelander-Institut für Astronomie, Universität Bonn, Auf dem Hügel 71, 53121, Bonn, Germany\\
  $^9$ Instituto de Astronom\'{i}a, Universidad Nacional Autonoma de Mexico, A. P. 70-264, C.P. 04510, Mexico, D.F., Mexico\\
  $^{10}$ Department of Physics and Astronomy, Oberlin College, Oberlin, Ohio, OH 44074, USA\\
  $^{11}$ INAF-Osservatorio di Astrofisica e Scienze dello Spazio di Bologna, Via Gobetti 93/3, I-40129 Bologna, Italy\\
  $^{12}$ Dipartimento di Fisica e Astronomia `Augusto Righi', Universita degli Studi di Bologna, Via Gobetti 93/2, I-40129 Bologna, Italy
}

\begin{document}

\maketitle

\begin{abstract}

  We present the extended ALMA MaNGA QUEnching and STar formation survey, a combination of the original 46 ALMaQUEST galaxies plus new ALMA observations for a further 20 interacting galaxies.  Three well-studied scaling relations are fit to the 19,999 star-forming spaxels in the extended sample, namely the resolved Schmidt-Kennicutt (rSK) relation, the resolved star forming main sequence (rSFMS) and the resolved molecular gas main sequence (rMGMS). We additionally investigate the relationship between the dynamical equilibrium pressure ($P_{DE}$) and star formation rate surface density (\sigsfr), which we refer to as the resolved $P_{DE}$ (rPDE) relation.  Contrary to previous studies that have focussed on normal star-forming galaxies and found an approximately linear rPDE relation, the presence of more vigourously star-forming galaxies in the extended ALMaQUEST sample reveals a marked turnover in the relation at high pressures.  Although the scatter around the linear fit to the rPDE relation is similar to the other three relations, a random forest analysis, which can extract non-linear dependences, finds that \textit{$P_{DE}$ is unambiguously more important than either \sigh2\ or \sigstar\ for predicting \sigsfr}.   We compare the observed rPDE relation to the prediction of the pressure-regulated feedback-modulated (PRFM) model of star formation, finding that galaxies residing on the global SFMS do indeed closely follow the rPDE relation predicted by the PRFM theory.  However, galaxies above and below the global SFMS show significant deviations from the model.  Galaxies with high SFR are instead consistent with models that include other contributions to  turbulence in addition to the local star formation feedback.
  
\end{abstract}

\begin{keywords}
Galaxies: evolution, galaxies: interactions, galaxies: starburst, galaxies: star formation, galaxies: ISM
\end{keywords}

\section{Introduction}\label{intro_sec}

Scaling relations provide a valuable tool for understanding the underlying drivers in physical processes.  In astronomy, correlations such as the Faber-Jackson relation (Faber \& Jackson 1976), the $M-\sigma$ relation (Ferrarese \& Merritt 2000) and the mass metallicity relation (Lequeux et al. 1979), have been used as both theoretical supports and interpretative tools for various aspects of galaxy evolution.  Examining the dependence on additional variables (e.g. Ellison et al. 2008b; Mannucci et al. 2010; Peng et al. 2010; Hunt et al. 2020; Morselli et al. 2020) provides further insight, with machine learning techniques now being employed to distinguish `fundamental' relations from mere correlations (e.g. Teimoorinia et al. 2016; Dey et al. 2019; Bluck et al. 2020; Baker \& Maiolino 2023).

In the field of star formation, an extensive literature has established a tight correlation between the surface densities of star formation rate and molecular gas, \sigsfr\ and \sigh2\ respectively (e.g. Wong \& Blitz 2002; Bigiel et al. 2008; Schruba et al. 2011), with possibly an even tighter relationship between star formation and the denser gas phase traced by HCN (Wu et al. 2005; Lada et al. 2010; Jimenez-Donaire et al. 2019).  Ostensibly, these scaling relations indicate that star formation is set by the abundance of available fuel (i.e. gas content).  However, the non-universality of star formation efficiencies on both global (e.g. Daddi et al. 2010; Saintonge et al. 2011) and local scales (e.g. Leroy et al. 2008; Usero et al. 2015; Ellison et al. 2021a; Thorp et al. 2022; Jimenez-Donaire et al. 2023) hints at additional physics that regulates star formation.

Early studies suggested that the fraction of the interstellar medium (ISM) in the molecular phase was driven by the internal pressure of the ISM, which then formed stars at an approximately constant efficiency  (e.g. Wong \& Blitz 2002; Blitz \& Rosolowsky 2004, 2006; Kim \& Ostriker 2007; Leroy et al. 2008).   These theories have been expanded in the last decade to describe star formation as a `self-regulating' process, that is set by the balance between energy/momentum input from massive stars and the weight of the ISM (e.g. Ostriker et al. 2010; Ostriker \& Shetty 2011; Shetty \& Ostriker 2012; Kim et al. 2013; Ostriker \& Kim 2022).  As a result of this balance, the SFR is predicted to scale approximately linearly with the gravity felt by the ISM.  Such models are referred to as pressure-regulated feedback-modulated (PRFM) theories.

Several contemporary observational studies have supported the PRFM formalism, that star formation is not only strongly correlated with the ISM pressure (e.g.  Herrera-Camus et al. 2017; Fisher et al. 2019; Sun et al. 2020a, 2023), but that relations between the SFR and combinations of both gas and stellar mass are tighter and stronger than the resolved star-forming main sequence (rSFMS), resolved Schmidt-Kennicutt (rSK) relation or resolved molecular gas main sequence (rMGMS) (e.g. Shi et al. 2011; Barrera-Ballesteros et al. 2021a,b).  Moreover, Barrera-Ballesteros et al. (2021a) found that the hydrostatic ISM pressure and \sigsfr\ correlate consistently over a wide range of galaxy morphologies, thus proposing it as the main parameter that regulates star formation on kpc scales.  Most recently, the data compilation and comparison with the latest generation of high resolution hydrodynamical simulations presented by Ostriker \& Kim (2022) show strong support for the PRFM model.  However, to date, the data that have been included in these comparisons have been drawn almost exclusively from local spiral galaxies on the star-forming main sequence. Unlike earlier star formation models (e.g., Tan 2000; Krumholz et al. 2018, Semenov et al. 2019), the PRFM formalism has not yet been tested against a wider range of galaxies that display differing star formation behaviour, such as starbursts (e.g., Daddi et al. 2010, Sharda et al. 2019), metal-poor dwarfs (e.g., Jameson et al. 2016), and early types that retain gas (e.g., Davis et al. 2014).  Here, we aim to further the comparison of the PRFM model with an additional dataset that includes a wider range of galaxies than those previously assessed.

The ALMA MaNGA QUEnching and STar formation (ALMaQUEST) survey aims to understand the kpc-scale processes that regulate star formation in the nearby universe. The original ALMaQUEST sample consists of 46 galaxies (Lin et al. 2020), spanning a wide range of SFRs from the green valley (e.g. Lin et al. 2022) to the starburst regime (Ellison et al. 2020a).   Although other molecular gas surveys have either higher spatial resolution such as  the Physics at High Resoution in Nearby Galaxies survey (PHANGS; Leroy et al. 2021), or a larger sample such as the Extragalactic Database for Galaxy Evolution (EDGE-CALIFA; Bolatto et al. 2017), ALMaQUEST's diversity of star-forming properties provides vital leverage for studying the mechanisms that both boost and quench star formation.  Moreover, we have recently enlarged the ALMaQUEST sample by almost 50\% by observing a further 20 galaxies in ALMA's Cycle 7 (PI Pan).  In keeping with the survey's goal of probing a wide range of environments, the new sample focuses on interacting galaxies, allowing us to study the relative role of fuel supply and star formation efficiency in boosting star formation (Thorp et al. 2022).  We will refer to the combined sample of 46 original ALMaQUEST galaxies plus the 20 mergers as the extended ALMaQUEST sample.

In this paper we have five main goals.   First, we re-visit the three scaling relations (rSK, rSFMS and rMGMS) that we have studied in our previous works, in order to present a complete assessment for the extended dataset.  Second, we provide a public release of the star formation rate, stellar mass and molecular gas surface densities for all star-forming spaxels in the extended sample of 66 galaxies to permit a reproduction of these relations. Third, we separate the sample into subsets of control galaxies that represent normal, face-on orientations that we might expect to manifest fiducial relations, then mergers and central starbursts which represent more extreme populations.  The star formation scaling relations are examined separately in each of these subsets.  Fourth, we investigate the relation between dynamical equilibrium pressure ($P_{DE}$) and \sigsfr\ (hereafter the resolved dynamical equilibrium pressure relation, rPDE) in ALMaQUEST galaxies for the first time, in order to assess whether (as suggested by Barrera-Ballesteros et al. 2021a) this relation might be the fundamental regulator of star formation.  Finally, we compare the observed rPDE relation with contemporary theories of star formation in order to assess whether (and under what conditions) the models can reproduce the data.

The paper is organized as follows.  In Section \ref{sample_sec} we describe the extended ALMaQUEST sample, as well as three sub-samples that are defined for the purposes of this paper.  In Section \ref{data_sec} we describe the MaNGA and ALMA data products that are released with this paper. The star formation scaling relations for the extended ALMaQUEST sample, with a particular focus on the rPDE relation, are presented in Section \ref{relations_sec} with a more detailed discussion of results in Section \ref{discussion_sec}.  Our conclusions are presented in Section \ref{conclusions_sec}.

\section{The extended ALMaQUEST sample}\label{sample_sec}

The extended ALMaQUEST sample consists of a total of 66 galaxies, all of which were drawn from the MaNGA DR15.   ALMA observations (described below) were obtained through five separate regular proposals. In ALMA Cycles 3, 5 and 6 proposals 2015.1.01225.S, 2017.1.01093.S, 2018.1.00558.S (PI: Lin) obtained CO(1-0) data for a sample of galaxies that spanned both the star-forming main sequence and the green valley, in order to primarily investigate quenching (Ellison et al. 2021b; Lin et al. 2022).  The Cycle 6 proposal 2018.1.00541.S (PI: Ellison) complemented this sample by focussing on galaxies with central starbursts, again with main sequence galaxies included as comparison targets, in order to investigate the physical processes that lead to enhanced star formation rates (Ellison et al. 2020a, b).  The combination of these four proposals includes 46 unique galaxies and represents the main ALMaQUEST sample that is described by Lin et al. (2020) and that has been used to investigate the scaling relations of star forming regions (Lin et al. 2019; Ellison et al. 2021a, Lin et al. 2022).  More recently, in ALMA Cycle 7 (2019.1.00260.S, PI Pan), we obtained CO(1-0) data for a further 20 MaNGA selected galaxies.  The Cycle 7 sample focused on interacting galaxies and contains targets that have close companions and/or show evidence for tidal features (Thorp et al. 2022).  Taken together, the final extended ALMaQUEST sample therefore spans a broad range of galactic properties that can be used to investigate a variety of environments and star formation regimes.

In addition to the complete sample of 66 galaxies, we here define three galaxy sub-samples that will be used in this paper's analysis.  First, we select a sample of galaxies that we expect to be minimally affected by issues such as inclination, starbursts or mergers.  This `control' sample consists of 16 galaxies which have axial ratios $b/a \ge 0.35$ and are not in either of the central starburst, nor merger samples (see below).  Second, we define a sample of central starburst galaxies.  Following Ellison et al. (2020a), central starburst galaxies are identified by plotting the radial profiles of \sigsfr\ (as determined from H$\alpha$ emission; see Section 3.1 for more details) and selecting galaxies with enhancements of at least 0.2 dex within the inner 0.5 $R/R_e$.  Although this selection process is identical to that of Ellison et al. (2020a), here we enforce the additional criterion that the galaxy not be in a merger, which excludes three of the central starbursts from the Ellison et al. (2020a) sample (8081-9101, 8156-3701, 8615-3703).  However, the Cycle 7 observations that have been obtained since the publication of Ellison et al. (2020a) include two galaxies that qualify as central starbursts and do not show merger features (8085-12701, 8085-3704). As a result of these choices, there are 11 central starburst galaxies in our sample.

Finally, we define a sample of galaxy mergers.  Although our Cycle 7 proposal focussed exclusively on mergers, there are also some interacting galaxies in the main (original) ALMaQUEST sample.  Moreover, the Cycle 7 sample includes galaxies that span a wide range of interaction stages, including pairs that do not show any tidal features and therefore may either yet to have experienced a pericentric passage, or those whose features are either fundamentally weaker (due to the orbital configuration) or fainter (due to fading).  We therefore elect to define our merger sample from scratch, by visually inspecting both the SDSS imaging, as well as the deeper Dark Energy Camera Legacy Survey (DECaLS) imaging that exists for the full sample.  Galaxies that exhibit clear morphological disturbances, such as bridges, tidal arms or shells, are included in the merger sample.  19 galaxies fulfill this requirement, the majority of which are post-mergers, i.e. single galaxies assumed to be observed after coalescence (e.g. Ellison et al. 2013; Thorp et al. 2019; Bickley et al. 2021, 2022), but some are still readily identifiable as interacting pairs (e.g. Thorp et al. 2022).   Although we did not allow mergers to be in the central starburst sample, we do allow mergers to have central starbursts (there are six such galaxies in the merger sample).  That is, all galaxies in Table \ref{sample_tab} identified as mergers are in the merger sample, but the galaxies in Table \ref{sample_tab} with central starbursts and also identified as mergers are not in our central starburst sample.  Images for each galaxy in the extended ALMaQUEST sample are available in either the original ALMaQUEST survey description paper (Lin et al. 2020) or in the presentation of the Cycle 7 data for the mergers (Thorp et al. 2022).

Table \ref{sample_tab} summarizes the targets in the extended ALMaQUEST sample and presents their integrated stellar masses and SFRs taken from the PIPE3D (S\'{a}nchez et al., 2016a, 2016b) value-added catalog (VAC, S\'{a}nchez et al. 2018), derived through summing individual spaxel values across the MaNGA data cubes and the integrated molecular gas masses described in Section \ref{alma_sec} from our ALMA data. The molecular gas masses are calculated assuming a conversion factor $\alpha_{CO}$ = 4.3 M$_{\odot}$ pc$^{-2}$ (K km s$^{-1}$)$^{-1}$ (including the contribution from helium).  In contrast with many of our previous studies that have used a Salpeter initial mass function (IMF), which is the default option adopted by PIPE3D, in the work presented here we have converted \sigsfr\ and \sigstar\ values to a Chabrier IMF in order to be more readily comparable to other work in the literature (e.g. Sun et al. 2020a; Barrera-Ballesteros et al. 2021a).  This conversion from Salpeter to Chabrier IMFs is achieved by multiplying the former by a factor of 0.61, see Madau \& Dickinson (2014).

\begin{table*}
  \caption{Summary of global (within the IFU) properties for the extended ALMaQUEST sample, taken from either the PIPE3D Value-added catalog, or derived from our ALMA observations.  The Proposal ID indicates the initial of the PI's surname (E=Ellison, L=Lin, P=Pan) and the ALMA Cycle in which the data were obtained.}
\begin{tabular}{lcccccccc}
\hline
Plate-IFU  &  $z$   &  log M$_{\star}$& log SFR & Log M(H$_2$) & Axial & Proposal  &   Merger?  &  Central   \\
 & & (VAC) & (VAC) &   (within IFU) & ratio &  ID &   & starburst? \\
  &    & (M$_{\odot}$) & (M$_{\odot}$ yr$^{-1}$) &  (M$_{\odot}$)&    &   &    &     \\
\hline                                                                                                      
 7977$-$3704     &     0.02724      &      10.14       &   $-$0.60      &    8.73        &       0.66       &       L3   &    0	 &     0 \\
 7977$-$12705    &     0.02724      &      10.64       &    0.25        &     9.41      &       0.67       &       L3   &    1	 &     0 \\
 7977$-$9102     &     0.06331      &      10.71       &    0.61        &    9.85       &       0.63       &       P7   &    1	 &     0 \\
 7977$-$3703     &     0.02781      &      10.14       &    0.18        &    9.16       &       0.80       &       E6   &    0	 &     1  \\
 7977$-$9101     &     0.02656      &      11.02       &   $-$0.08      &    8.79        &       0.55       &       L3   &    1	 &     0 \\
 7975$-$6104     &     0.07920      &      10.80       &    0.40        &    9.96       &       0.88       &       P7   &    1	 &     0 \\
 7968$-$12705    &     0.08638      &      11.08       &   $-$1.63      &    $<$7.13        &       0.47       &       P7   &    1     &     0 \\
 8153$-$12702    &     0.03823      &       9.68       &   $-$0.95      &    $<$7.06        &       0.47       &       P7   &    1	 &     0 \\
 8077$-$9101     &     0.04323      &      10.20       &   $-$0.31      &    8.90        &       0.18       &       L6   &    0     &     0 \\
 8082$-$12701    &     0.02703      &      10.27       &   $-$0.09      &     8.90       &       0.62       &       E6   &    0	 &     0 \\
 8081$-$3704     &     0.05400      &      10.34       &    0.75        &    9.26       &       0.96       &       E6   &    0	 &     1  \\
 8082$-$9102     &     0.03652      &      10.53       &    0.51        &    9.82       &       0.31       &       P7   &    1	 &     1 \\
 8085$-$12701    &     0.02980      &      10.22       &    0.37        &     8.71      &       0.52       &       P7   &    0	 &     1 \\
 8081$-$9101     &     0.02846      &      10.39       &    0.11        &    9.45       &       0.55       &       L5   &    1	 &     1  \\
 8083$-$12702    &     0.02104      &      11.01       &    0.45        &     9.55      &       0.74       &       L5   &    0	 &     0 \\
 8085$-$3704     &     0.03707      &      10.51       &    0.44        &    9.74       &       0.52       &       P7   &    0	 &     1 \\
 8156$-$3701     &     0.05272      &      10.31       &    0.66        &    9.02       &       0.73       &       E6   &    1	 &     1 \\
 8086$-$9101     &     0.04003      &      10.73       &   $-$0.10      &    9.20        &       0.50       &       L5   &    0	 &     0 \\
 8078$-$6103     &     0.02859      &      10.53       &    0.40        &    9.62       &       0.59       &       L5   &    0	 &     0 \\
 8081$-$6102     &     0.03719      &      10.58       &   $-$0.61      &    8.72        &       0.54       &       L5   &    0     &     0 \\
 8081$-$12703    &     0.02558      &      10.12       &   $-$1.00      &     8.93       &       0.20       &       L5   &    0	 &     0 \\
 8084$-$3702     &     0.02206      &      10.01       &    0.21        &    9.30       &       0.66       &       E6   &    0	 &     0 \\	
 8085$-$6101     &     0.05180      &      10.85       &   $-$0.95      &    $<$7.56         &       0.64       &       P7   &    1	 &     0 \\
 8078$-$12701    &     0.02698      &      10.73       &    0.19        &     9.56      &       0.38       &       L5   &    0	 &     0 \\
 8078$-$6104     &     0.04449      &      10.20       &    0.02        &    9.18       &       0.44       &       P7   &    1	 &     0 \\
 8081$-$9102     &     0.03407      &      10.45       &    0.03        &    9.28       &       0.17       &       L5   &    0	 &     0 \\
 8155$-$6102     &     0.03081      &      10.14       &    0.14        &    9.27       &       0.89       &       E6   &    0	 &     0 \\
 8241$-$12705    &     0.02719      &      10.18       &  $-$0.21       &     9.20       &       0.49       &       P7   &    1     &     0 \\
 8241$-$3703     &     0.02911      &       9.89       &    0.04        &    8.83       &       0.94       &       E6   &    0	 &     0 \\
 8241$-$3704     &     0.06617      &      10.79       &    1.00        &    10.03      &       0.95       &       E6   &    0	 &     1   \\
 8083$-$6101     &     0.02677      &      10.09       &  $-$1.09       &    9.32        &       0.23       &       L5   &    0	 &     0 \\
 8084$-$12705    &     0.02545      &      10.24       &  $-$0.24       &     8.85       &       0.22       &       L5   &    0	 &     0 \\
 8155$-$6101     &     0.03740      &      10.71       &  $-$0.64       &    9.11        &       0.74       &       L5   &    0	 &     0 \\
 8077$-$6104     &     0.04601      &      10.52       &    0.43        &    9.53       &       0.93       &       E6   &    0	 &     0 \\
 8082$-$12704    &     0.13214      &      11.20       &    0.55        &     10.21     &       0.44       &       L5   &    0	 &     0 \\
 8083$-$9101     &     0.03847      &      10.92       &    0.38        &    9.65       &       0.33       &       L6   &    1     &     0 \\
 8078$-$12703    &     0.02832      &      10.60       &  $-$0.08       &     9.12      &       0.77       &       P7   &    0	 &     0 \\
 8084$-$6103     &     0.03593      &      10.30       &  $-$0.35       &    9.40        &       0.31       &       L5   &    0	 &     0 \\
 8083$-$12703    &     0.02466      &      10.24       &    0.10        &     9.04      &       0.30       &       P7   &    1     &     1 \\
 8082$-$6103     &     0.02416      &      10.09       &    0.17        &    8.94       &       0.74       &       E6   &    0	 &     1  \\
 8615$-$12702    &     0.02095      &       9.99       &  $-$0.19       &     8.55       &       0.27       &       E6   &    0	 &     0 \\
 8623$-$6104     &     0.09704      &      11.07       &    0.96        &    10.12      &       0.47       &       E6   &    0	 &     0 \\
 8616$-$9102     &     0.03039      &      10.23       &    0.44        &    9.42       &       0.82       &       E6   &    0	 &     1  \\
 8615$-$3703     &     0.01845      &       9.98       &    0.19        &    9.28       &       0.55       &       E6   &    1	 &     1  \\
 8952$-$6104     &     0.02843      &      10.12       &    0.24        &    9.11       &       0.93       &       L5   &    0	 &     0 \\
 8950$-$12705    &     0.02528      &      10.32       &  $-$0.62       &     9.44       &       0.15       &       L5   &    0	 &     0 \\
 8655$-$3701     &     0.07149      &      10.94       &    0.90        &    10.43      &       0.55       &       L6   &    0	 &     0 \\	
 9194$-$3702     &     0.07454      &      10.84       &    0.75        &    10.32      &       0.82       &       P7   &    1     &     0 \\
 9195$-$3703     &     0.02724      &      10.18       &    0.26        &    9.32       &       0.56       &       P7   &    1     &     0 \\
 8728$-$3701     &     0.02833      &      10.42       &  $-$0.59       &    8.70        &       0.73       &       L5   &    1     &     0 \\
 7815$-$12705    &     0.02955      &      10.62       &    0.20        &     9.72      &       0.34       &       L5   &    0	 &     1  \\
 8616$-$12702    &     0.03083      &      10.54       &  $-$0.45       &     8.41       &       0.78       &       L5   &    1	 &     0 \\
 8615$-$9101     &     0.03346      &      10.37       &  $-$0.18       &    9.12        &       0.26       &       L5   &    0	 &     0 \\
 8616$-$9101     &     0.09197      &      10.88       &    0.53        &    9.81       &       0.50       &       P7   &    1	 &     0 \\
 8952$-$12701    &     0.02856      &      10.52       &  $-$0.60       &     8.91       &       0.60       &       L5   &    1	 &     0 \\ 
 8450$-$6102     &     0.04200      &      10.21       &    0.43        &     9.23  &       0.76       &       E6   &    0	 &     1  \\
\hline
\end{tabular}
\label{sample_tab}
\end{table*}

\addtocounter{table}{-1}
\begin{table*}
  \caption{Continued}
\begin{tabular}{lcccccccc}
\hline
Plate-IFU  &  $z$  &   log M$_{\star}$& log SFR & Log M(H$_2$) & Axial & Proposal  &   Merger?  &  Central   \\
 & & (VAC) & (VAC) &   (within IFU) & ratio &  ID &   & starburst? \\
  &    & (M$_{\odot}$) & (M$_{\odot}$ yr$^{-1}$) &  (M$_{\odot}$)&  &   &    &     \\
\hline 
 9195$-$3702     &     0.06434     &     10.93      &    0.89     &     9.98    &    0.61       &       P7   &    1     &     0 \\
 9195$-$9101     &     0.05687     &     10.53      &    0.17     &     9.45    &    0.81       &       P7   &    1	 &     0 \\
 8618$-$9102     &     0.04334     &     10.24      &    0.21     &     9.27    &    0.33       &       L6   &    0	 &     1  \\
 8623$-$1902     &     0.02495     &      9.95      &  $-$1.22    &     8.30     &    0.92       &       P7   &    1     &     0 \\
 8615$-$1901     &     0.02019     &      9.47      &  $-$0.29    &     $<$7.30      &    0.85       &       P7   &    1	 &     1 \\
 8623$-$12702    &     0.02691     &     10.49      &  $-$0.57    &      9.49    &    0.17       &       L5   &    0	 &     0 \\
 8616$-$6104     &     0.05426     &     10.56      &    0.04     &     9.37    &    0.45       &       L5   &    0	 &     0 \\
 8655$-$12705    &     0.04557     &     10.12      &  $-$0.93    &      8.78     &    0.49       &       L5   &    1	 &     0 \\
 8655$-$9102     &     0.04509     &     10.20      &    0.06     &     8.82    &    0.71       &       E6   &    0	 &     1   \\
 9512$-$3704     &     0.05463     &     10.48      &  $-$0.07    &     8.96    &    0.68       &       P7   &    1	 &     0 \\
\hline
\end{tabular}
\end{table*}

\section{Data}\label{data_sec}

Since the acquisition and processing of ALMaQUEST data have been described extensively in our previous works (e.g. Lin et al. 2019, 2020; Ellison et al. 2020a, b, 2021a, b), we give only a brief overview of salient details here and refer readers requiring more detail to the aforementioned works.

\subsection{MaNGA data}\label{manga_sec}

We use MaNGA data from the Data Release 15 as the source of all resolved quantities such as emission line fluxes and stellar mass surface densities.  Spaxels are 0.5 arcseconds on a side and are thus over-sampled compared with both the fibre size (2 arcseconds) and typical seeing.  Although the analysis presented here (and in all of our previous papers) uses the 0.5 arcsecond spaxels, we have repeated all of our analysis smoothing to a range of angular scales (up to 3 arcseconds) and find no qualitative difference in our results.

Optical emission line fluxes are taken from the public PIPE3D data cubes (S\'{a}nchez et al. 2016a,b, 2018) and corrected for internal extinction by assuming an intrinsic H$\alpha$/H$\beta$=2.85 and a Milky Way extinction curve (Cardelli, Clayton \& Mathis 1989).   Star formation rate surface densities (\sigsfr) are computed from H$\alpha$ luminosities using Equation 2 from Kennicutt (1998), a technique that has been shown to reproduce the UV and IR SFRs well in integral field unit (IFU) data (Catalan-Torrecilla et al. 2015). We discuss the potential limitations of these \sigsfr\ measurements for the work presented here in more detail in Section 5.3.5. Surface densities of stellar mass (\sigstar) are also taken from the PIPE3D catalog.  All surface density quantities are inclination corrected using the axial ratio ($b/a$) listed in Table \ref{sample_tab}.

In the next section, we will investigate the scaling relations of star-forming spaxels.  To qualify as star-forming, any given spaxel must meet three conditions.   First, we require that the MaNGA spaxel has a S/N$>$2 in each of the four optical emission lines used in the Baldwin, Phillips \& Terlevich (1981; hereafter BPT) diagram, i.e., H$\alpha$, H$\beta$, [OIII]$\lambda$5007, [NII]$\lambda$6584.   In practice, this means that the strongest line (H$\alpha$) is observed with a much higher S/N, but this does not affect the results of this work.  Second, the emission line ratios of the spaxel must lie below the criterion defined by Kauffmann et al. (2003), thus identifying it as being dominated by stellar photoionization.  Third, we impose an H$\alpha$ equivalent width (EW) cut H$\alpha>6$ \AA\ (e.g. Cid-Fernandes et al. 2011) to remove potential `retired' spaxels (which have been studied explicitly for the ALMaQUEST sample by Ellison et al. 2021b and Lin et al. 2022).     

\subsection{ALMA data}\label{alma_sec}

The acquisition and processing of ALMA data for the main sample of 46 galaxies is described in detail in Lin et al. (2020). The extended ALMaQUEST sample, including the 20 new galaxies observed in Cycle 7, follows identical procedures (described in more detail in Thorp et al. 2022).  Below we review the details relevant for the current work.

CO(1-0) (hereafter, simply CO) spectral line observations were obtained between 2016-2020 in the array's second most compact configuration (C43-2).  The single pointing primary beam size for this configuration is $\sim$ 50 arcsec with an angular resolution $\sim$ 2.5 arcsec.  Integration times ranged from 0.2 to 2.5 hours on source, using one high resolution spectral window focused on the CO line and one to three additional low resolution continuum windows for calibration.  The data cubes were all processed using the Common Astronomy Software Applications (CASA; McMullin et al. 2007) package.  The final cubes have channel widths of 11 \kms\ and root mean square (RMS) noise of $\sigma_{rms}$ = 0.2 -- 2 mJy beam$^{-1}$. To permit a mapping of the ALMA data cubes onto the MaNGA data products, the ALMA data were first trimmed to the size of that galaxy's MaNGA cube (MaNGA IFU bundles range in size from 12 -- 32 arcsec chosen to match the galaxy size).  A fixed restoring beam size of 2.5 arcsec with pixel size of 0.5 arcsec was then applied to the ALMA cube.  These two steps resulted in ALMA data cubes with the same size and sampling as the MaNGA data products.  

CO luminosities in each pixel (L'(CO) in Jy \kms\ pc$^2$ pix$^{-1}$) were converted to molecular gas surface densities (\sigh2) using a conversion factor ($\alpha_{CO}$) such that \sigh2\ (M$_{\odot}$ pix$^{-1}$) = $\alpha_{CO} \times $ L'(CO).  In Table \ref{spaxel_tab} we provide \sigh2\ both per pixel and per kpc$^{2}$.   In keeping with our previous ALMaQUEST papers, we assume a fixed conversion factor $\alpha_{CO}$ = 4.3 M$_{\odot}$ pc$^{-2}$ (K km s$^{-1}$)$^{-1}$ (including the contribution from helium) which is a typical value adopted for the Milky Way (e.g. Bolatto et al. 2013).  As with other surface densities used in this work, \sigh2\ is corrected for inclination using the $b/a$ axial ratio given in Table \ref{sample_sec}.  Typical uncertainties in \sigh2\ are $<$ 0.1 dex.

In Table \ref{sample_tab} we report the integrated molecular gas mass within the MaNGA IFU footprint (assuming our fiducial $\alpha_{CO}$=4.35 M$_{\odot}$ pc$^{-2}$ (K km s$^{-1}$)$^{-1}$ ) that can be used to compute a `global' gas fraction or star formation efficiency by comparing with the VAC stellar mass or star formation rate.  Four galaxies in the Cycle 7 sample are not detected in the integrated maps.  In these cases the 3$\sigma$ upper limit of the CO flux is calculated as $3\sigma_{RMS}\times{\sqrt{\delta v \Delta V}}$, where $\sigma_{RMS}$ is the RMS noise from the spectral line data cube, $\delta v$~$=$~11~km~s$^{-1}$ is the velocity resolution, and $\Delta V$ is the assumed 500~km~s$^{-1}$ width.

There is a total of 19,999 spaxels that fulfill both the star-forming criteria described in Section \ref{manga_sec} and have detections in CO with a S/N$>$3, that we henceforth refer to as the `full' star-forming spaxel sample.   These spaxels have all of the necessary measurements required to investigate the various star formation scaling relations.   The control, central starburst and merger samples contain 8321, 5357 and 4176 star-forming spaxels respectively (these numbers do not sum to 19,999 because some galaxies appear in more than one sub-sample, and some galaxies appear in none).  We note that one of the control galaxies and two of the merger galaxies have no star-forming spaxels.  The \sigsfr, \sigstar\ and \sigh2\ values for the full sample of (CO-detected) 19,999 star-forming spaxels are listed in Table \ref{spaxel_tab}.

There exist numerous alternative prescriptions for variable conversion factors, taking into account parameters such as metallicity and SFR (e.g. Narayanan et al. 2012; Bolatto 2013; Sandstrom et al. 2013; Accurso et al. 2017; Hunt et al. 2020; Gong et al. 2020).  As an alternative to using a fixed conversion factor, we therefore also compute \sigh2\ with the metallicity dependent formalism of Sun et al. (2020b):

\begin{equation}\label{alpha_eqn}
  \alpha_{CO,Z} = 4.35 Z^{-1.6} M_{\odot} pc^{-2} (K~km~s^{-1})^{-1}.
\end{equation}

In order to determine $\alpha_{CO,Z}$ from Equation \ref{alpha_eqn}, an accurate gas phase metallicity must be available for the spaxel.  We therefore only compute $\alpha_{CO,Z}$ (and the subsequent values of \sigh2) when the emission lines of H$\alpha$, H$\beta$, [OIII]$\lambda$5007, [NII]$\lambda$6584 all have S/N$>5$ and the spaxel lies below the Kauffmann et al. (2003) line that separates star-forming and AGN dominated zones.  Metallicities are computed using the O3N2 calibration of Pettini \& Pagel (2004) normalized to a solar value of 12+log(O/H)=8.69.  Since the S/N requirement for accurate metallicity determination (S/N$>$5) is stricter than for our nominal star-forming sample (S/N$>$2) of the full sample of 19,999 star-forming spaxels, only 16,254 have metallicity measurements (with values mostly in the range 8.55 $<$ 12 + log O/H $<$ 8.75) and hence have metallicity dependent determinations of \sigh2.  The impact of the choice of conversion factor on our work (as well as tests with other possible values) is discussed extensively in Section \ref{discussion_sec}.

\begin{table*}
  \caption{Spaxel properties for all CO detected (S/N$\ge$3) star-forming spaxels in the extended ALMaQUEST sample.  Spaxel X and Y coordinates are given as unitless quantities (starting at 0,0) to enable a reconstruction of the 2 dimensional maps. All surface density quantities are inclination corrected. The first 10 rows are given here as an example; the full dataset is available from the online journal.}
\begin{tabular}{lccccccc}
\hline
Plate-IFU  &  X, Y  & Log L'(CO)                      & CO   & Log \sigh2\    & Log \sigh2\     & Log \sigstar\               & Log \sigsfr\  \\
           &       & (Jy \kms\ pc$^2$ pix$^{-1}$)  & S/N  & (M$_{\odot}$ pix$^{-1}$) & (M$_{\odot}$ kpc$^{-2}$)     & (M$_{\odot}$ kpc$^{-2}$) & (M$_{\odot}$ yr$^{-1}$ kpc$^{-2}$)   \\
\hline                                                                                                      

7977$-$3704  & 28,10   &   5.27   &   3.21  &    5.90   &    6.85   &    8.08  &   -2.35 \\ 
7977$-$3704  & 29,10   &   5.27   &   3.17  &    5.90   &    6.84   &    8.08  &   -2.37  \\
7977$-$3704  & 29,11   &   5.40   &   4.32  &    6.03   &    6.98   &    8.08  &   -2.29  \\
7977$-$3704  & 30,11   &   5.36   &   3.92  &    5.99   &    6.94   &    8.08  &   -2.44  \\
7977$-$3704  & 31,11   &   5.27   &   3.20  &    5.90   &    6.85   &    7.88  &   -2.67  \\
7977$-$3704  & 26,12   &   5.31   &   3.51  &    5.94   &    6.89   &    8.25  &   -2.15  \\
7977$-$3704  & 27,12   &   5.39   &   4.24  &    6.03   &    6.97   &    8.25  &   -2.15  \\
7977$-$3704  & 29,12   &   5.46   &   4.96  &    6.09   &    7.04   &    8.12  &   -2.21  \\
7977$-$3704  & 30,12   &   5.43   &   4.64  &    6.06   &    7.01   &    8.12  &   -2.34  \\
7977$-$3704  & 31,12   &   5.37   &   3.99  &    6.00   &    6.95   &    7.88  &   -2.56  \\

\hline
\end{tabular}
\label{spaxel_tab}
\end{table*}

\section{Star formation scaling relations}\label{relations_sec}

\begin{figure*}
	\includegraphics[width=8cm]{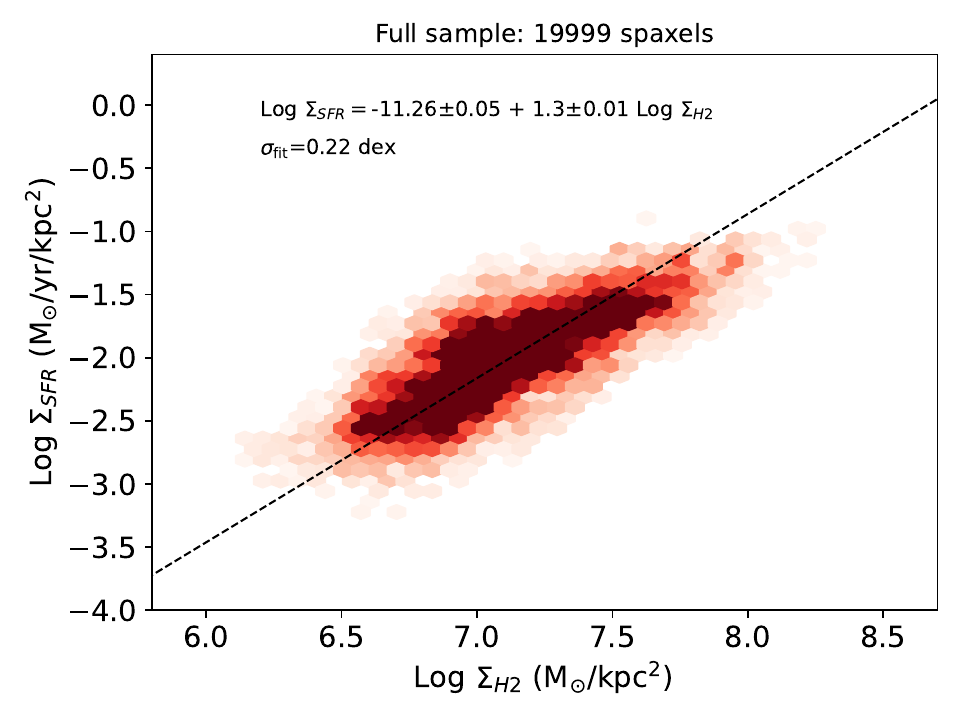}
	\includegraphics[width=8cm]{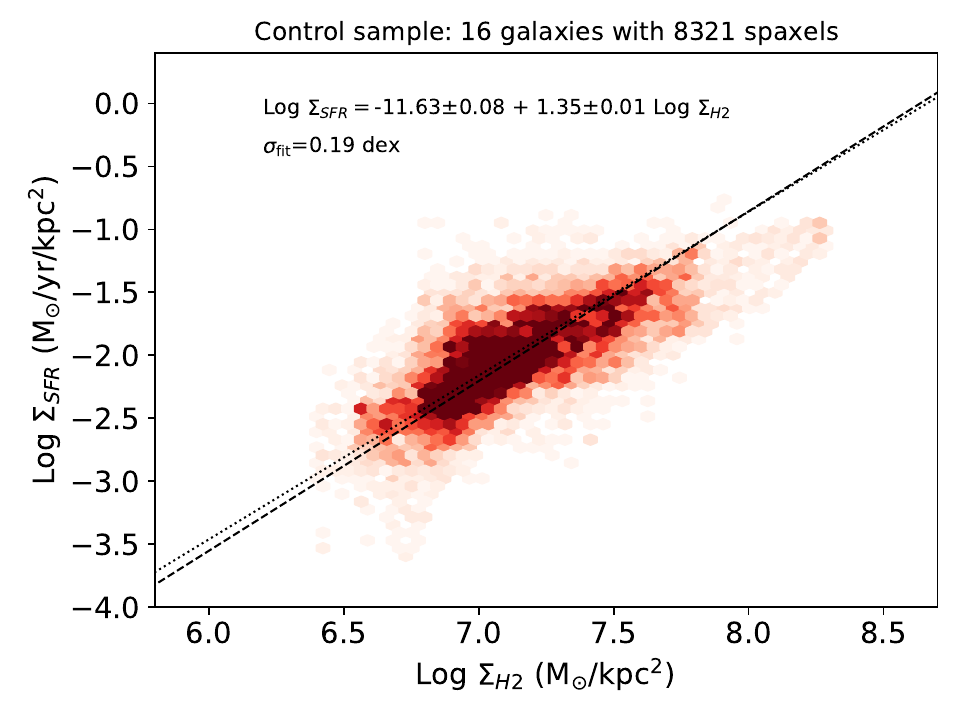}
	\includegraphics[width=8cm]{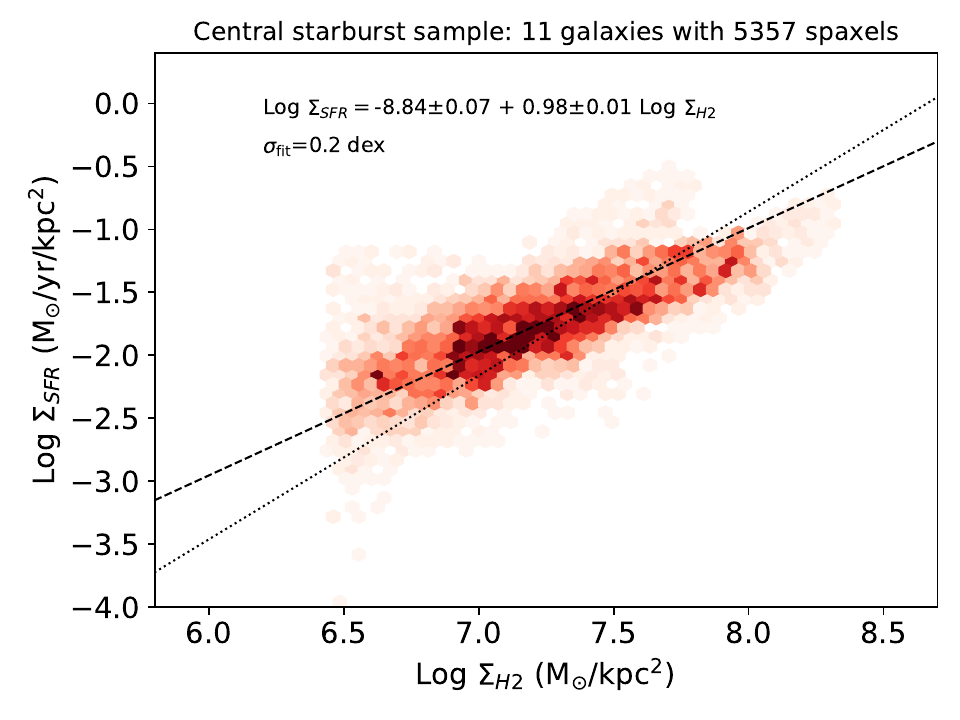}
	\includegraphics[width=8cm]{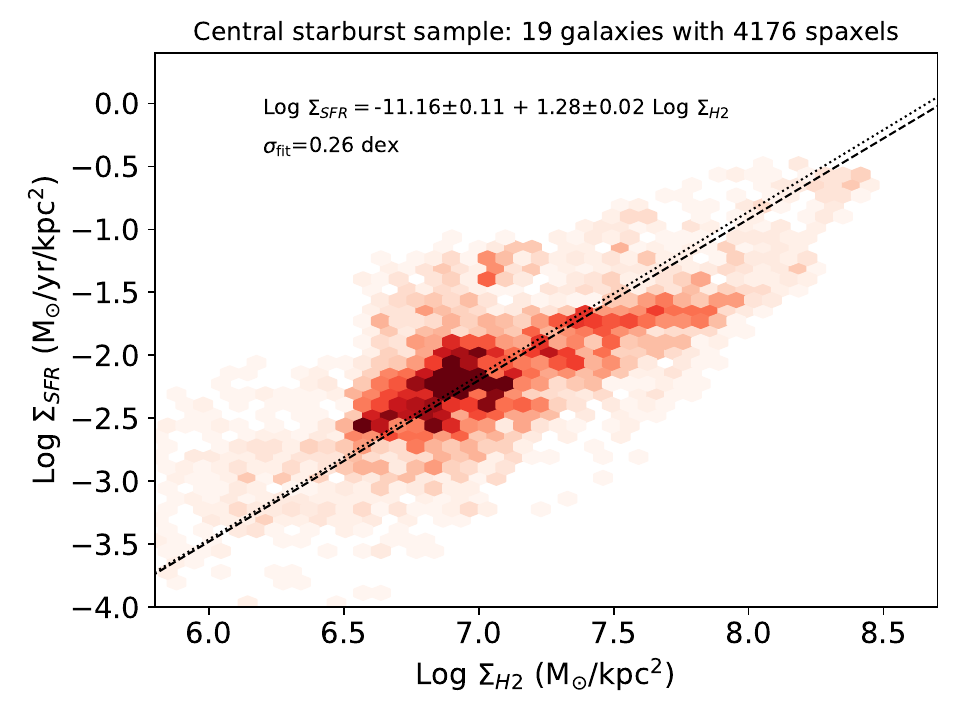}
        \caption{The resolved Schmidt-Kennicutt relation for the four galaxy samples defined in this work.  In each panel the ODR fit is shown by the dashed line.  The dotted line reproduces the fit for the full star-forming sample as a reference.}
    \label{ks_fig}
\end{figure*}

\begin{figure*}
	\includegraphics[width=8cm]{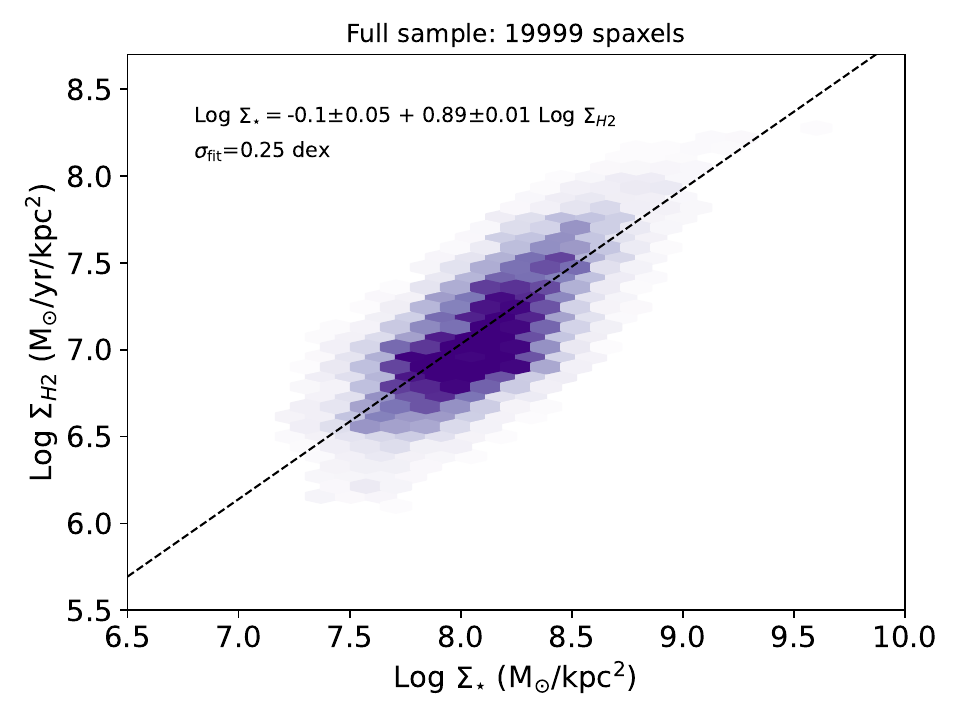}
	\includegraphics[width=8cm]{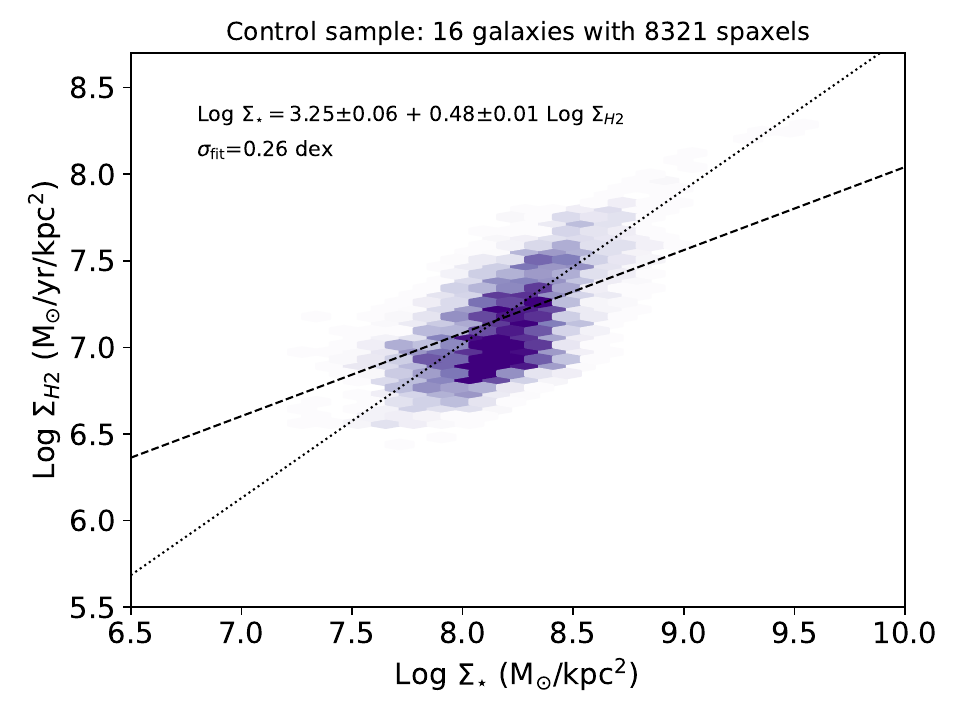}
	\includegraphics[width=8cm]{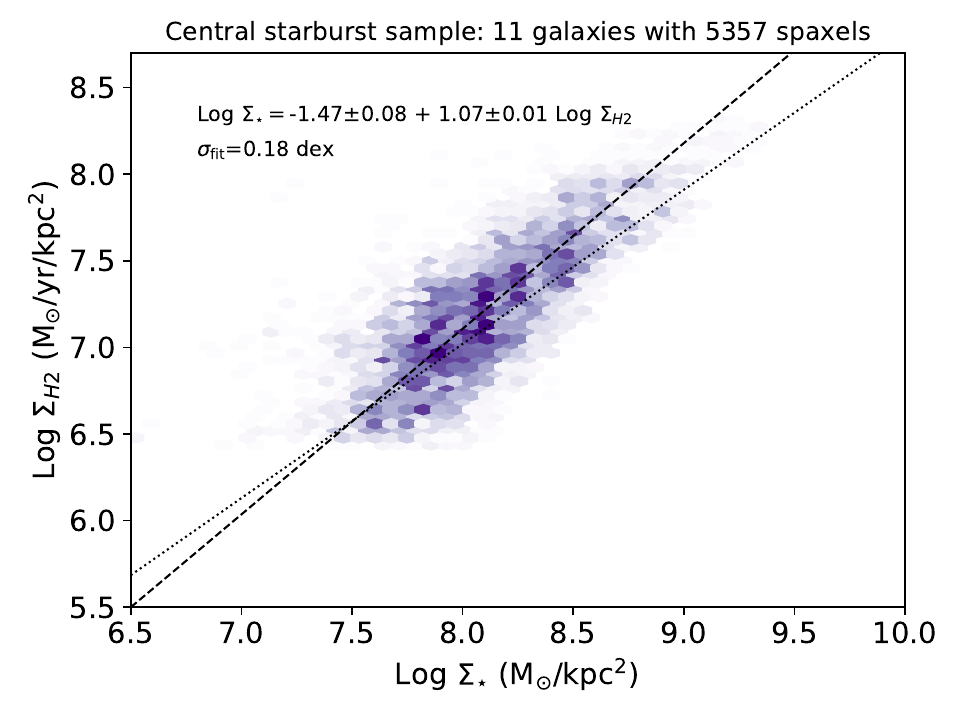}
	\includegraphics[width=8cm]{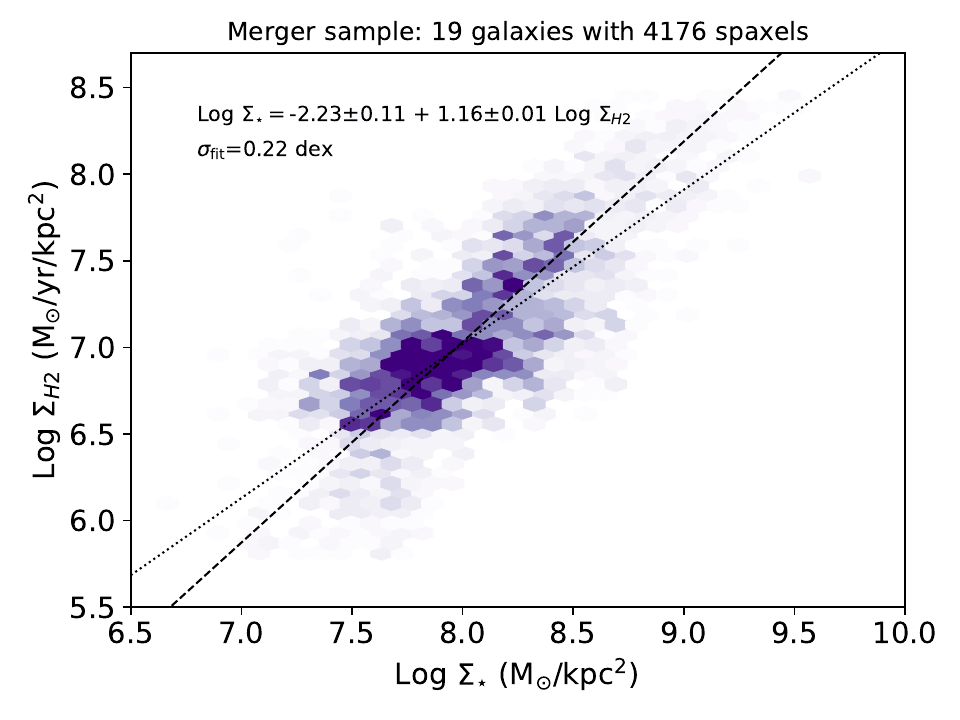}
        \caption{The resolved molecular gas main sequence for the four galaxy samples defined in this work.  In each panel the ODR fit is shown by the dashed line.  The dotted line reproduces the fit for the full star-forming sample as a reference.}
    \label{mgms_fig}
\end{figure*}

One of the main focuses of this paper is to present the rPDE relation for star-forming spaxels in the ALMaQUEST survey, and to assess whether a) it is universal (i.e. invariant between galaxies) and b) whether it is more fundamental than the previously studied rSK, rMGMS and rSFMS relations.  Although these latter three scaling relations have already been presented in several previous ALMaQUEST papers (e.g. Lin et al. 2019; Ellison et al. 2021a; Lin et al. 2022), it is useful to re-visit them here for several reasons.  First, we are presenting the extended ALMaQUEST sample, including 20 new galaxies to complement the original 46, for the first time.  Second, in the previous section we introduced three sub-samples of galaxies (controls, central starbursts and mergers) to capture the diversity in the dataset.  Before investigating the rPDE relation for these samples, as well as for the full extended ALMaQUEST sample, we therefore briefly present the rSK, rMGMS and rSFMS relations for context and comparison.

\subsection{The Resolved Schmidt-Kennicutt Relation}

\begin{figure*}
	\includegraphics[width=8cm]{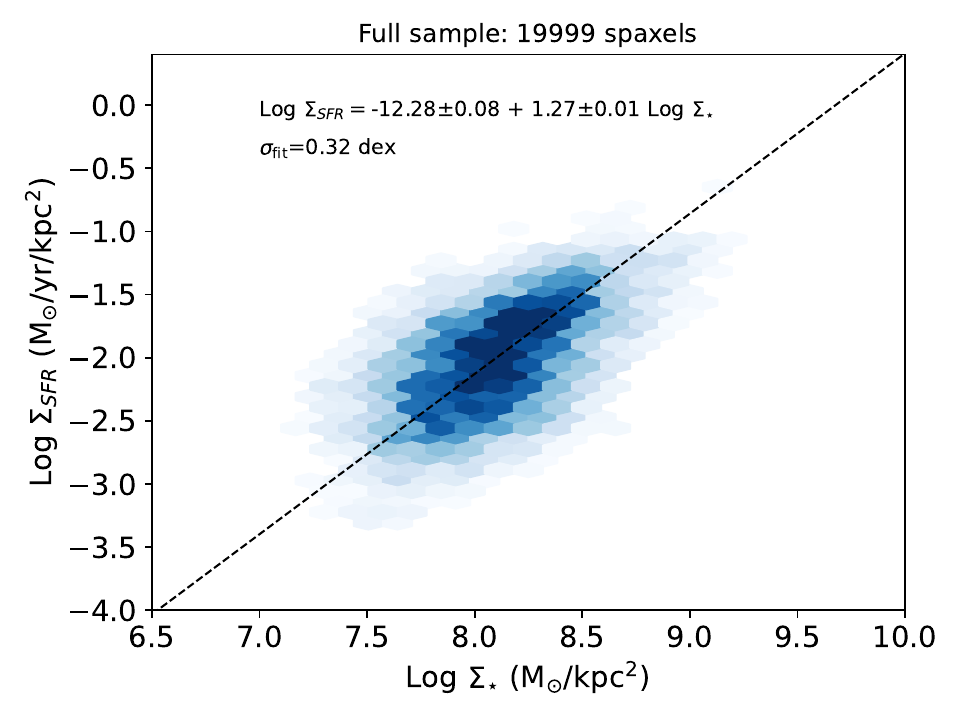}
	\includegraphics[width=8cm]{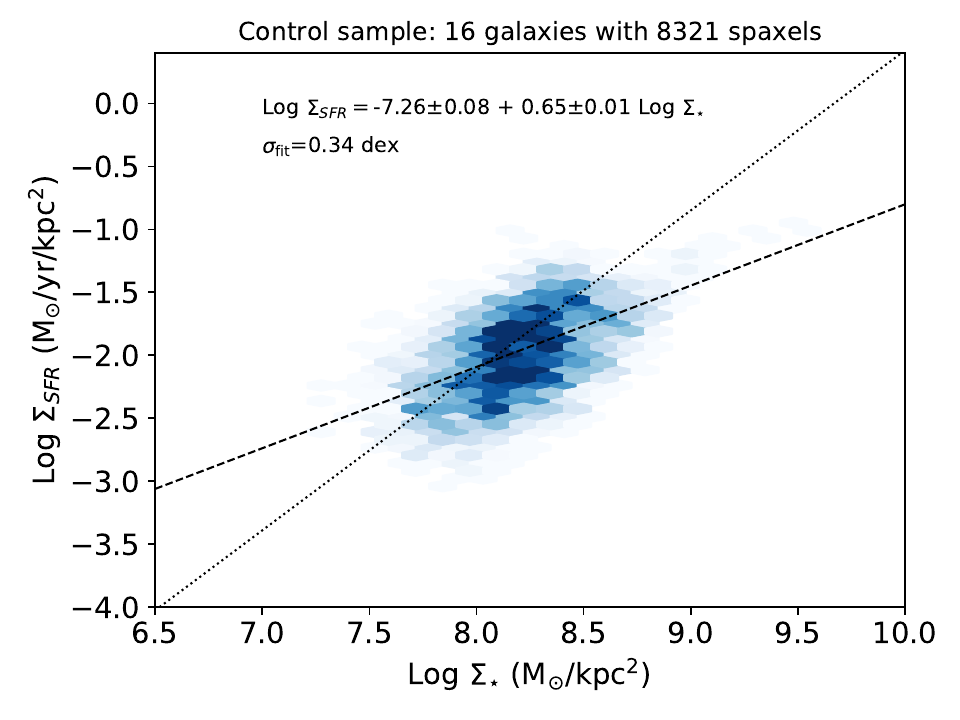}
	\includegraphics[width=8cm]{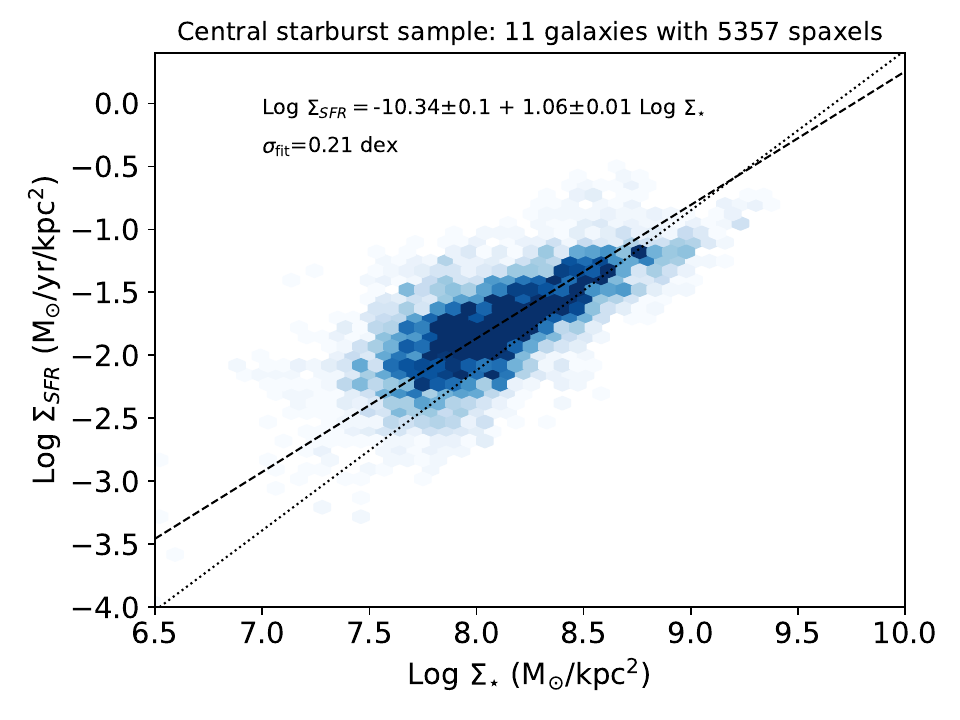}
	\includegraphics[width=8cm]{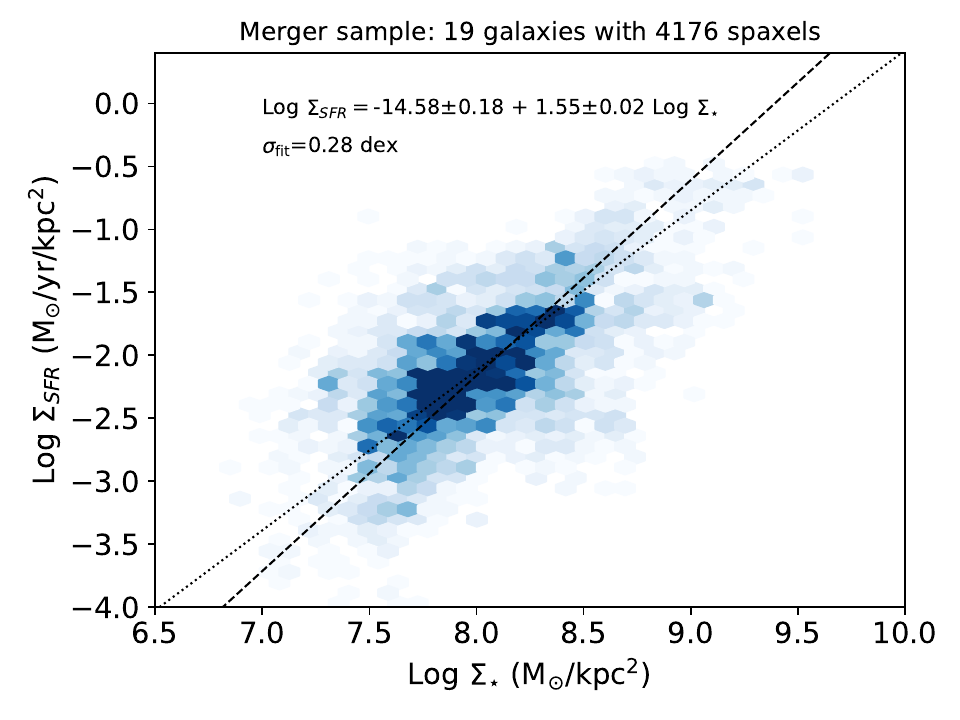}
        \caption{The resolved star forming main sequence for the four galaxy samples defined in this work.  In each panel the ODR fit is shown by the dashed line.  The dotted line reproduces the fit for the full star-forming sample as a reference.}
    \label{sfms_fig}
\end{figure*}

In the top left panel of Figure \ref{ks_fig} we present the rSK relation for all 19,999 star-forming spaxels in the extended ALMaQUEST sample of 66 galaxies.  In the remaining three panels we show the rSK relation for the other galaxy sub-samples considered in this work: the `control' sample of non-merger, non-central starburst galaxies with moderate inclinations in the top right panel, as well as the mergers and central starbursts in the bottom two panels. The dashed line in each panel shows the orthogonal distance regression (ODR) fit, with fit coefficients given explicitly in each panel.  The dotted line in the top right and bottom two panels reproduces the ODR fit to the full sample (i.e. the dashed line in the top left panel) and is shown for reference.  A strong rSK exists for all of these galaxy sub-samples with scatter of 0.2--0.3 dex, with the smallest scatter measured in the control galaxies and in the central starbursts, and the largest in the mergers.  Thorp et al. (2022) presented a dedicated study of mergers in ALMaQUEST (although their sample is defined slightly differently from ours and includes some pre-coalescence galaxy pairs).  One of their conclusions was that some galaxy mergers have high SFRs driven by an elevated star formation efficiency (SFE = \sigsfr / \sigh2), which would naturally increase the scatter in the rSK relation (see also Garay-Solis et al. 2023 for a study of the central molecular gas content in CALIFA-selected mergers).  However, elevated SFEs were also found by Ellison et al. (2020a) in the sample of ALMaQUEST central starbursts, and yet the rSK relation remains relatively tight in the lower left panel of Figure \ref{ks_fig}, indicating that deviations are typically smaller in this sample.  The central starbursts therefore have broadly self-similar SFEs (even if they are enhanced), whereas Thorp et al. (2022) found that, in mergers, enhanced gas fractions could also lead to SFR enhancements.

\subsection{The Resolved Molecular Gas Main Sequence}

In Figure \ref{mgms_fig} we show the rMGMS for the four galaxy sub-samples; once again, a strong relationship is seen for each one with the central starbursts exhibiting the tightest relationship.  Indeed, the rMGMS for the central strburst sample has the smallest scatter of all of the relations and samples studied in this work.  Therefore, despite central starbursts being selected to be deviant (in their central star formation), their gas fractions are apparently quite self-similar.  In contrast to the rSK relation, the rMGMS for the control sample exhibits the largest scatter, and the ODR fit yields a flatter slope than for the other samples.  The precise details of the relation in a given sample is likely to be driven by variations in the individual galaxies (e.g. Ellison et al. 2021a; Pessa et al. 2022).  Overall though, we again find a scatter that is 0.2 - 0.3 dex across the various sub-samples, consistent with values in previous works, not only for ALMaQUEST (Lin et al. 2019; Ellison et al. 2021a), but also for other galaxy samples studied at kpc-scales (e.g. Pessa et al. 2021; S\'{a}nchez et al. 2021; Casasola et al. 2022). 

\subsection{The Resolved Star Forming Main Sequence}

Figure \ref{sfms_fig} completes the trio of the standard star formation scaling relations by showing the rSFMS.  As found in previous studies (e.g. Lin et al. 2019; Morselli et al. 2020; Pessa et al. 2021) the scatter is larger than the previous two relations, up to 0.34 dex.  Indeed, it has been suggested that the rSFMS (and, by extension, its global counterpart) is simply a by-product of other physical correlations (e.g. Lin et al. 2019; Ellison et al. 2021a; Baker et al. 2022, 2023).  Once again though, we note the relative tightness of the rSFMS exhibited by the central starbursts, only 0.21 dex in the ODR fit.  We note also the wide range of best fit slopes obtained for the rSFMS amongst the different samples, ranging between 0.65 and 1.55.   These results serve to underline the caution required when comparing the scaling relations between different works; even when a consistent fitting method is used, the details of the galaxy sample can significantly impact the best fit relation.

\subsection{The Resolved Dynamical Equilibrium Pressure Relation}

\begin{figure*}
	\includegraphics[width=8cm]{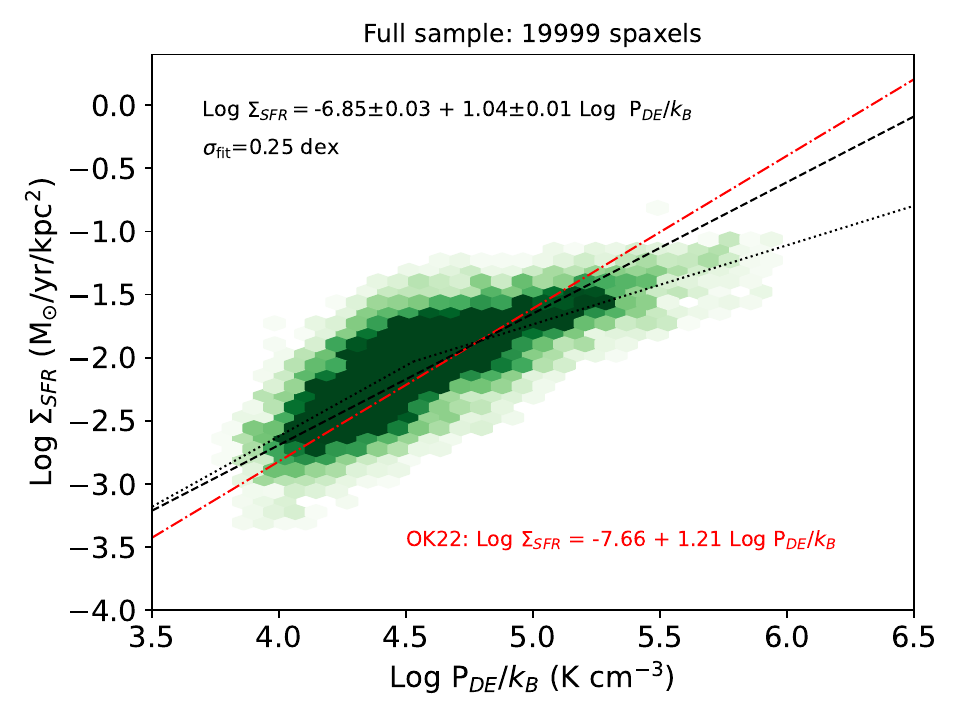}
	\includegraphics[width=8cm]{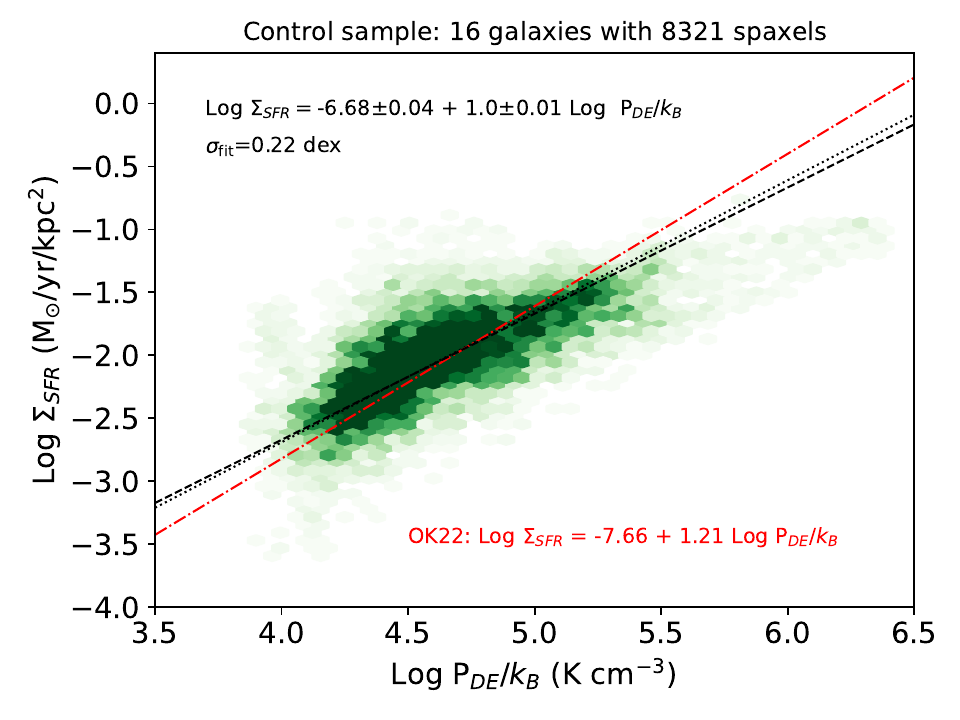}
	\includegraphics[width=8cm]{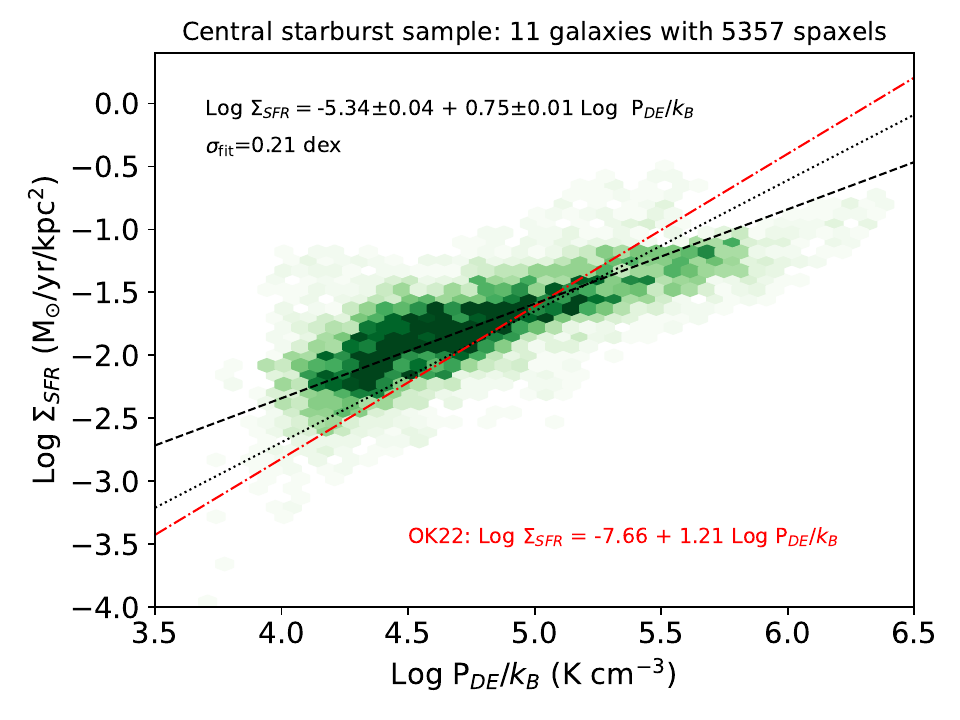}
	\includegraphics[width=8cm]{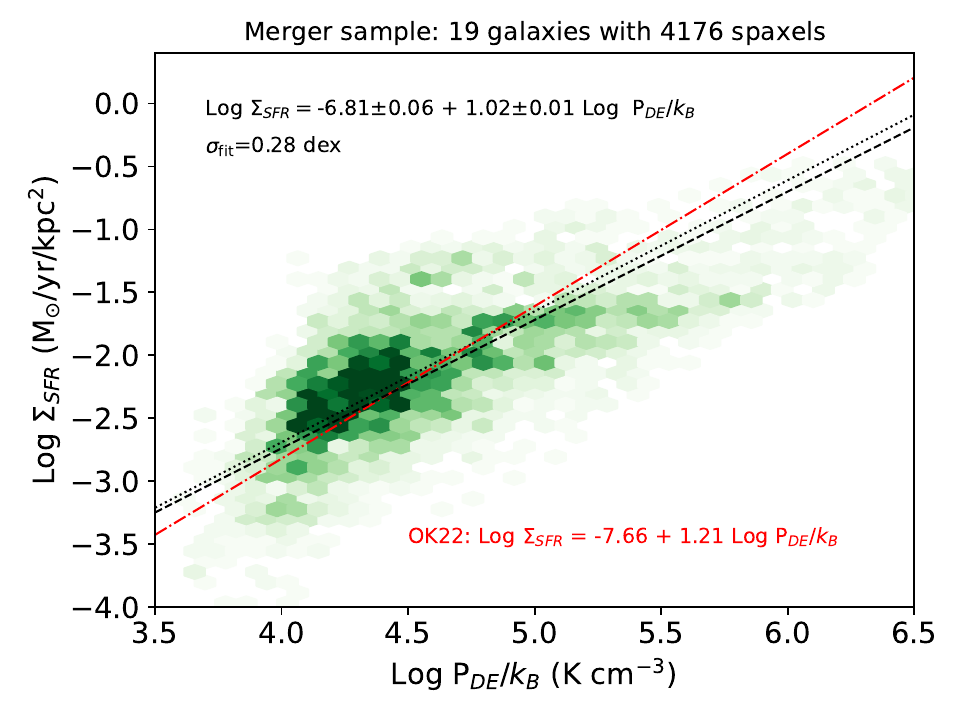}
        \caption{The resolved dynamical equilibrium pressure relation for the four galaxy samples defined in this work.  In each panel the ODR fit is shown by the dashed line. The red dot-dashed line shows the fit to the TIGRESS simulation of Ostriker \& Kim (2022).  The dotted line in the upper left panel shows the broken power law fit described in Equations \ref{2lawa} and \ref{2lawb}.  In the other three panels the dotted line shows the single power law fit to the full star forming spaxel sample.}
    \label{prfm_fig}
\end{figure*}

Having reviewed the three principal star formation scaling relations previously discussed in the literature and their characteristics in the ALMaQUEST sample, we now turn to the rPDE relation which has been assessed in both the PHANGS (Sun et al. 2020a, 2023) and EDGE-CALIFA surveys (Barrera-Ballesteros et al. 2021a), as well as in the nearby KINGFISH sample (Herrera-Camus et al. 2017).  Each of these surveys has their own benefits and limitations.  PHANGS has an order of magnitude better physical resolution than either EDGE-CALIFA or ALMaQUEST, but companion IFU data exists for only 19 galaxies (e.g. Groves et al. 2023).  The EDGE-CALIFA sample is the largest of the three (126 galaxies in the original sample; Bolatto et al. 2017), but is limited almost exclusively to galaxies close to the SFMS.  The KINGFISH sample had additional [CII] measurements that allowed a complementary analysis of the thermal pressure.  The niche of ALMaQUEST is its dynamic range in galactic properties, both in terms of a dedicated merger component, as well as populations of both starbursts and green valley galaxies.

ISM pressure can be evaluated in numerous different ways, but recent papers have favoured $P_{DE}$ which evaluates the mid-plane dynamical equilibrium pressure in the galactic disk accounting for contributions from both gas and stars.  The general framework for computing the kpc-scale dynamical equilibrium pressure is well established in the literature (e.g. Ostriker et al. 2010; Kim et al. 2011; Gallagher et al. 2018; Fisher et al. 2019; Schruba et al. 2019), where the gas and stars in the galactic disk are modelled as isothermal fluids in a plane-parallel geometry.    Specifically, we follow the implementation of Sun et al. (2020a) who express $P_{DE}$ as

\begin{equation}\label{prfm_eqn}
  P_{DE} = \frac{\pi G}{2} \Sigma_{gas}^2 + \Sigma_{gas} \sqrt{2G\rho_{\star}} \sigma_{gas,z}.
\end{equation}

The first term represents the weight of the ISM due to self gravity and the second term is the weight of the ISM due to stellar gravity.  $\Sigma_{gas}$ is the total gas surface density, i.e. the sum of the atomic and molecular components, where $\Sigma_{gas}$ = \sigh2\ + $\Sigma_{HI}$. $\rho_{\star}$ and $\sigma_{gas,z}$ are the mid-plane stellar mass volume density and the gas velocity dispersion perpendicular to the disk, respectively.  All three of these terms require some assumptions for our dataset.

In the absence of a measurement of $\Sigma_{HI}$, we assume a fixed value of $\Sigma_{HI}$ = 7 M$_{\odot}$ pc$^{-2}$, which is a typical value in galactic disks (e.g. Bigiel et al. 2008).  For most of the spaxels in our sample the molecular gas surface density is considerably higher than this (e.g. Figure \ref{ks_fig}), so that the precise value of $\Sigma_{HI}$ for our sample is not expected to have a significant impact on our result.  Indeed, even using a value as large as  $\Sigma_{HI}$ = 9 M$_{\odot}$ pc$^{-2}$ (which is the value at which the atomic gas surface density is found to saturate in nearby star-forming galaxies; Bigiel et al. 2008) does not strongly impact our results.  Nonetheless, we return to the assumption of a fixed $\Sigma_{HI}$ in the Discussion.

In the absence of a robust measurement of the vertical gas velocity disperson, we follow Barrera-Ballesteros et al. (2021a) and assume a fixed value of $\sigma_{gas,z}$ = 11 \kms\ (see also Blitz \& Rosolowsky 2004, 2006; Leroy et al. 2008; Ostriker et al. 2010) which is a fairly universal value found for normal star-forming galaxies in the local universe (Kennicutt \& Evans 2012; Caldu-Primo et al. 2013).  The assumption of a fixed $\sigma_{gas,z}$ was assessed in Sun et al. (2020a) to lead to a small over-estimate in $P_{DE}$, but with a scatter that was generally within 0.2 dex.  Again, we will return to the assumption of a fixed $\sigma_{gas,z}$ in the Discussion.

In order to estimate $\rho_{\star}$, we again follow the previous works of Blitz \& Rosolowsky (2006), Leroy et al. (2008), Ostriker et al. (2010) and Sun et al. (2020a), and calculate the mid-plane stellar mass density as:

\begin{equation}\label{rho_eqn}
  \rho_{\star} = \frac{\Sigma_{\star}}{4 H_{\star}}  = \frac{\Sigma_{\star}}{0.54 R_{\star}}
\end{equation}

\noindent where $R_{\star}$ and $H_{\star}$ are the disk scale length and height respectively.   The first step in Equation \ref{rho_eqn} assumes an isothermal density profile in the vertical direction.  The second step assumes a fixed disk flattening of $R_{\star}/H_{\star}$=7.3.  We determine $R_{\star}$ by taking the half-light (effective) radius ($R_{50}$) from the NASA Sloan Atlas (NSA) and then convert to the scale length using

\begin{equation}
  R_{\star} = R_{50}/1.68.
\end{equation}

The above formalism, although widely adopted in previous literature, ignores locally enhanced gravity due to small scale structures.  Clumpy sub-structure contributes additional weight to the mid-plane pressure compared to the assumption (used here) of a smooth disk.  This effect is well demonstrated in the analysis of PHANGS data at various resolutions, where Sun et al. (2020a) find that the over-pressurization of the disk (as inferred by comparing $P_{DE}$ to the turbulent pressure) depends on the physical resolution at which the terms are evaluated.   Since molecular gas is expected to be clumpy on scales below the resolution of our data, the impact of sub-structure may lead to an under-estimate of $P_{DE}$ in the ALMaQUEST data.

With these caveats and assumptions in mind (which we will return to with a more extensive discussion in Section 5.3), in Figure \ref{prfm_fig} we present the rPDE relation for the extended ALMaQUEST sample.  The figure format follows that of Figures \ref{ks_fig} -- \ref{sfms_fig}, with all star-forming spaxels presented in the top left panel and spaxels in the control, central starburst and merger samples in the remaining three panels.  In addition to the ODR fit (black dashed line), we also show (red dot-dashed line) the theoretical prediction of the PRFM model of Ostriker \& Kim (2022), derived by fitting to the output of the high resolution TIGRESS simulation (their Equation 26c).

Figure \ref{prfm_fig} demonstrates that there exists a tight relationship between $P_{DE}$ and \sigsfr\ in the ALMaQUEST data with a scatter whose magnitude of $\sim$ 0.2 -- 0.3 dex is broadly consistent with that of the other three star formation scaling relations.  The rPDE relation also shares the same general characteristic of exhibiting the greatest scatter in the merger sample and the least in the central starburst sample.  This is perhaps not surprising, since the calculation of $P_{DE}$ derives from \sigstar\ and \sigh2\ that appear individually in the rKS relation, the rMGMS and the rSFMS.  We will return to a quantitative comparison of the four relations in the next section.

\begin{figure}
	\includegraphics[width=9cm]{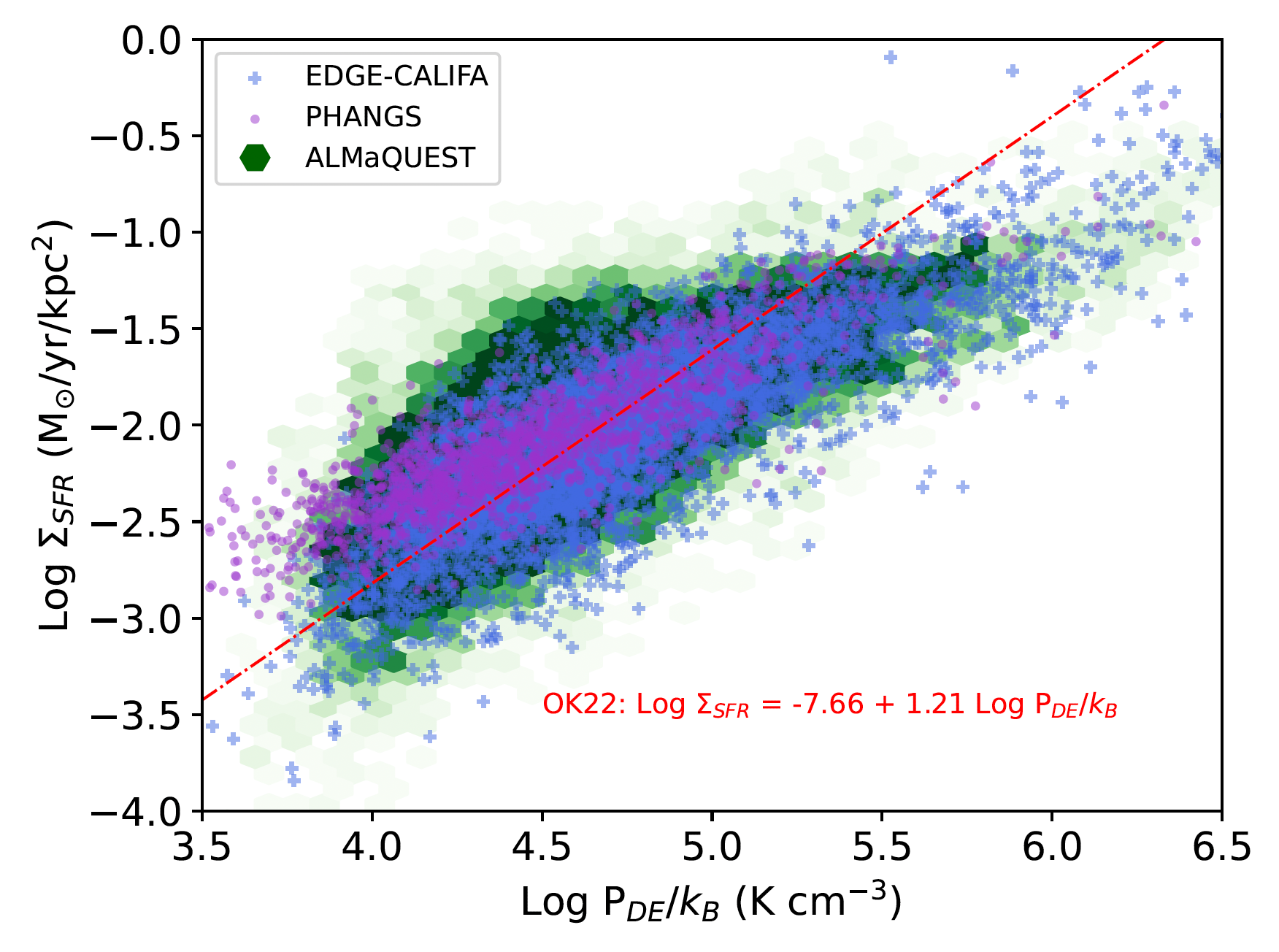}
        \caption{The rPDE relation for all star-forming spaxels in the extended ALMaQUEST dataset (the same data as shown in the top left panel of Figure \ref{prfm_fig}; green 2D histogram) shown with data from the EDGE-CALIFA survey (Barrera-Ballesteros et al. 2021a; blue crosses) and PHANGS (Sun et al. 2020; purple points).  The Ostriker \& Kim (2022) PRFM theory prediction is shown with the red dot-dashed line.  All three observational datasets are in broad agreement with one another and all of them fall systematically below the PRFM model at high pressures.  }
    \label{lit_fig}
\end{figure}

In terms of comparison to the PRFM formalism of Ostriker \& Kim (2022), Figure \ref{prfm_fig} shows that there is generally broad agreement with data; the red dot-dashed line (representing the model) typically passes through the data and (with the exception of the central starburst sample, which we return to below) the best fit relation is in reasonable agreement with the theoretical one.

However, a closer scrutiny of Figure \ref{prfm_fig} reveals some disparities between the PRFM model and the ALMaQUEST data.  For example, for the complete sample of 19,999 star-forming spaxels (top left panel of Figure \ref{prfm_fig}) it can be seen that the data do not follow a simple linear relation.  Instead, there is a break at log $P_{DE}/k_B > 4.5$ K cm$^{-3}$ beyond which the relationship flattens.  Such a turnover is not predicted by the Ostriker \& Kim (2022) model despite the fact that the TIGRESS simulations include the full range of pressures and star formation rates sampled by our data.  The flattening at high $P_{DE}$ can also be seen in the control sample (top right panel of Figure \ref{prfm_fig}).  Finally, the central starbursts also show a poorer match to the PRFM model; at low values of $P_{DE}$ the star formation rate surface densities are systematically higher than predicted by the model.  Conversely, at high $P_{DE}$, \sigsfr\  seems to fall short of the theory.  These two effects lead to a rPDE relation for the central starbursts that is significantly flatter than predicted by Ostriker \& Kim (2022).  In Section 5.3.4 we discuss some of the possible caveats in our measurements, although we will ultimately fail to identify any obvious culprit in the treatment of the data, or calculation of $P_{DE}$ that lead to this systematic disagreement.

Given the apparent turnover in the rPDE relation at log $P_{DE}/k_B \sim 4.5$ K cm$^{-3}$ we experimented with fitting a broken power law to the full star-forming spaxel sample, leaving both the exponents and the break point as free parameters.  The resulting fits are

\begin{equation}\label{2lawa}
    \log \Sigma_{SFR} = -7.09 + 1.12 \log (P_{DE} / k_B)
  \end{equation}

\noindent and

\begin{equation}\label{2lawb}
  \log \Sigma_{SFR} = -4.87 + 0.63 \log (P_{DE} / k_B)
  \end{equation}

\noindent for values of log $P_{DE}/k_B$ below and above 4.528 K cm$^{-3}$ respectively.  The broken power law is shown as a dotted line in the top left panel of Figure \ref{prfm_fig}.  Despite the visual improvement when using a double power law,  the residuals of the best fit broken power law are only 0.01 dex smaller than the single power law fit, so statistically a two component description does not offer an improvement over our original fit.  However, in the next section, we look more closely at the data that contribute to the rPDE relation in different regimes, which reveals distinct behaviour that depends on SFR.

\section{Discussion}\label{discussion_sec}

Building on earlier work (Ostriker et al. 2010; Ostriker \& Shetty 2011; Shetty \& Ostriker 2012; Kim et al. 2013), Ostriker \& Kim (2022) have presented the pressure-regulated feedback-modulated theory of star formation as a framework to relate both the availability of gas, as well as its physical state, to the production of stars.  By running a series of high resolution, multi-phase, magnetohydrodynamic simulations, Ostriker \& Kim (2022) predict that dynamical equilibrium pressure will scale with an approximately unity slope with the surface density of SFR.  The predicted scaling relation (referred to herein as the rPDE relation) showed good agreement with several previous observational studies (e.g. Leroy et al. 2008; Barrera-Ballesteros et al. 2021a; Sun et al. 2020a).  Moreover, Barrera-Ballesteros et al. (2021a) found that the relationship between hydrostatic pressure (a slight modification to the dynamical equilibrium pressure, but one that can be considered broadly equivalent) and \sigsfr\ was largely robust to variations in galaxy properties (e.g. morphology) in the EDGE-CALIFA sample.  The data presented in these papers support the PRFM model's fundamental tenet of a balance between star formation-driven feedback and ISM pressure that is widely applicable in local star-forming galaxies.

In contrast to these previous works, Figure \ref{prfm_fig} shows that, for the more diverse galaxy sample in the extended ALMaQUEST sample, the PRFM model is an incomplete representation of the dataset as a whole, and particularly so for the central starbursts.  Hints of a similar departure between theory and observations, which manifests as values of low \sigsfr\ at high $P_{DE}$, can in fact be glimpsed in other datasets.  For example, the rPDE relation in the EDGE-CALIFA dataset, although dominated by spaxels with lower $P_{DE}$ values than in ALMaQUEST, does in fact show a mild curvature (e.g Fig. 9 in Barrera-Ballesteros et al. 2021a and Fig. 15 in Ostriker \& Kim 2022).  This is demonstrated in Figure \ref{lit_fig} where we overplot the EDGE-CALIFA dataset (blue crosses; Barrera-Ballesteros et al. 2021a) as well as the PHANGS data (purple points, taken from Sun et al. 2020a) on top of the full set of star-forming spaxels in the extended ALMaQUEST sample (green scale).  Although these datasets have different selection functions and sample the gas at different physical scales, it can be seen that, over the majority of the range of pressures sampled by ALMaQUEST, all three surveys are in broad agreement with one another.   It can also be seen that, although neither the PHANGS nor EDGE-CALIFA datasets have extensive sampling of the high $P_{DE}$ regime, these samples fall under the Ostriker \& Kim (2022) prediction in the same way as seen for ALMaQUEST when log $P_{DE}/k_B > 5.0 $ K cm$^{-3}$.  Significant deviations from both the theoretical prediction and low $z$ field galaxy samples have also been reported by Fisher et al. (2019) who studied highly turbulent star-forming galaxies in the DYNAMO sample.  Taken together, these results raise the question of how universal the rPDE relation (and by extension, its underpinning PRFM theory) is.  In the following sub-sections we discuss this point, as well as possible caveats to our data.

\begin{figure}
	\includegraphics[width=8.5cm]{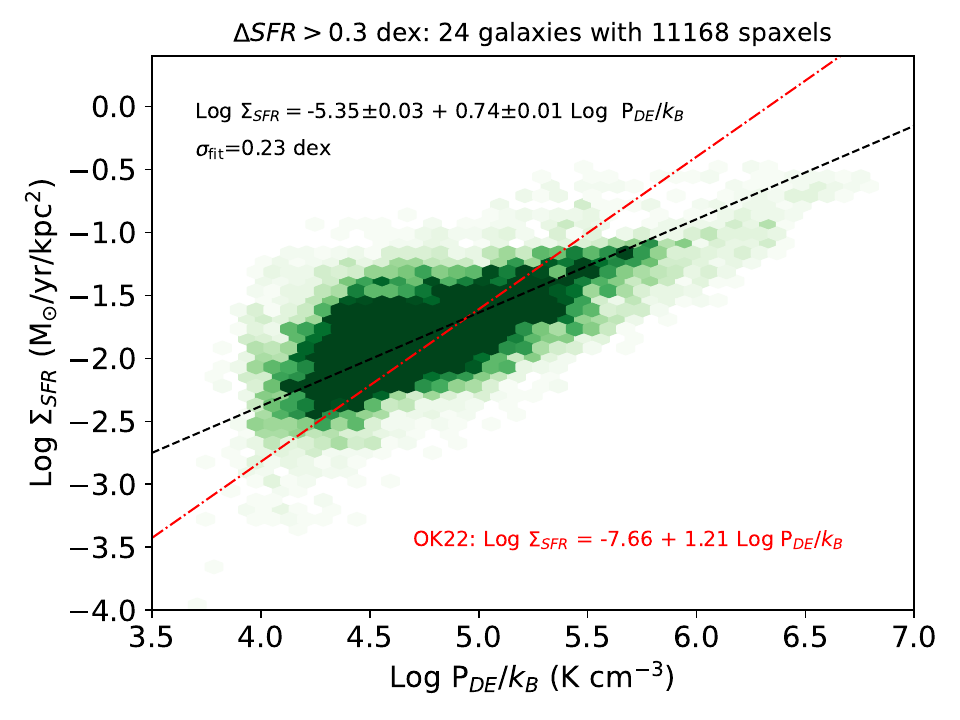}
	\includegraphics[width=8.5cm]{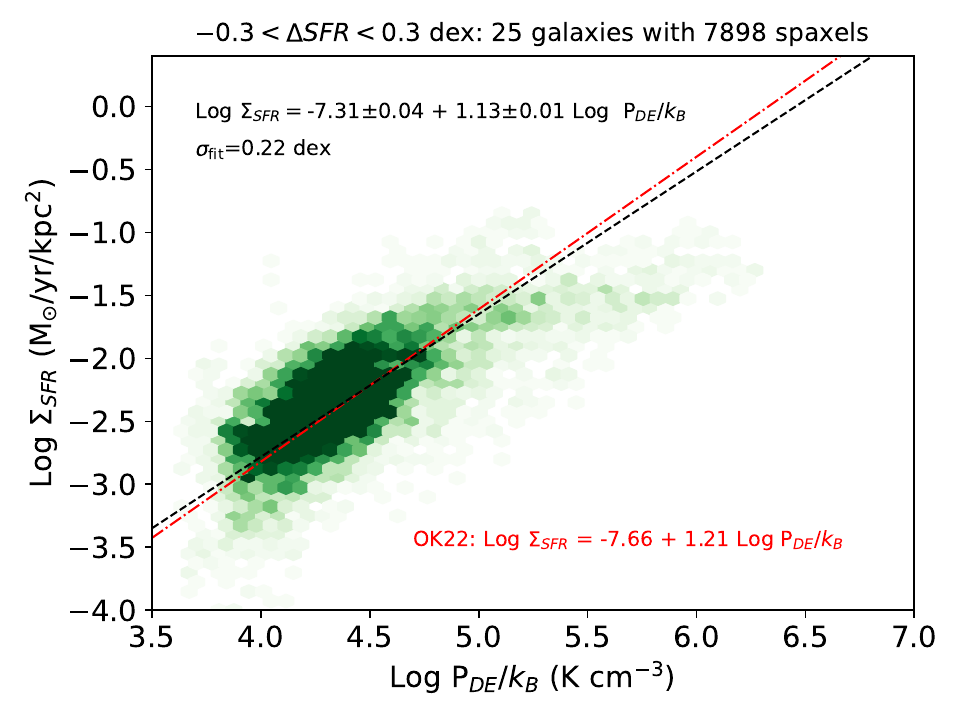}
	\includegraphics[width=8.5cm]{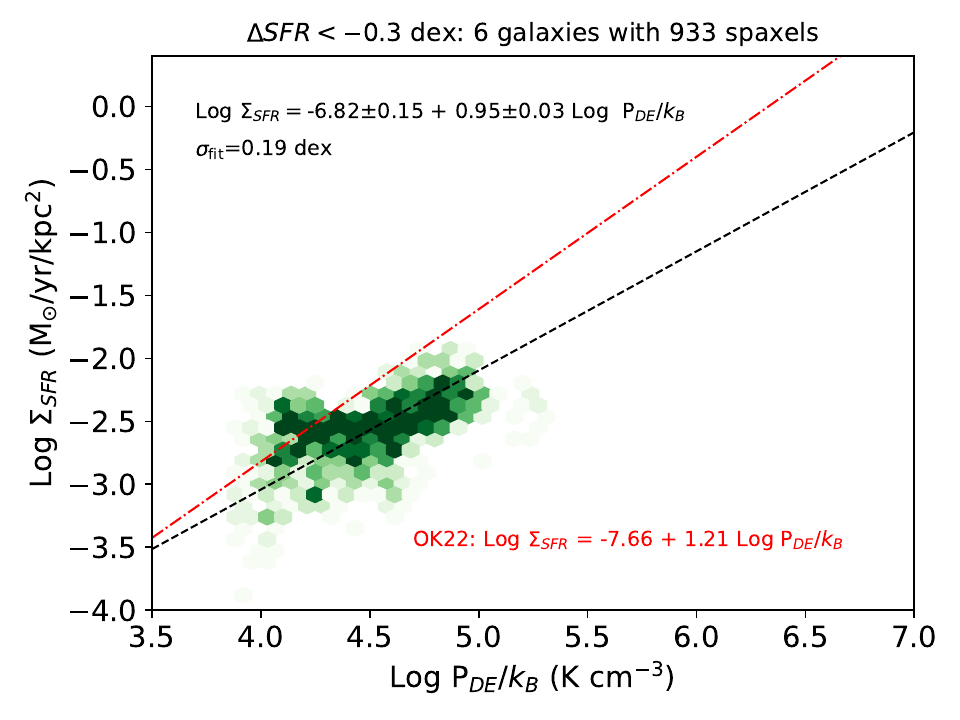}
        \caption{The rPDE relation for galaxies in three bins of $\Delta$ SFR.  The red dot-dashed line shows the prediction from the Ostriker \& Kim (2022) PRFM model.  Panels from bottom to top show star-forming spaxels located in galaxies with a global SFR that is at least a factor of two below the SFMS, within a factor of two of the SFMS, or at least a factor of two above the SFMS.}
    \label{pde_dsfr}
\end{figure}

\subsection{Is the rPDE relation universal?}

In order to further investigate whether (as indicated by the overall flattened rPDE relation seen in central starbursts) it is the presence of highly star-forming galaxies in our sample that leads to a departure from the theoretical expectation of the PRFM formalism of Ostriker \& Kim (2022), we separate the extended ALMaQUEST sample into three bins of SFR.  Specifically, we calculate an offset from the global SFMS (i.e. using total stellar masses and star formation rates) on a galaxy-by-galaxy basis.  This is achieved by comparing a given galaxy's SFR to a control sample matched within 0.1 dex in total stellar mass, within 0.1 dex of 5th nearest neighbour density and within 0.005 in redshift, such that $\Delta$SFR captures (in log units) the excess, or deficit, of star formation compared to the `norm' for a given stellar mass, environment and redshift (see Ellison et al. 2018 for more details).

In Figure \ref{pde_dsfr} we show the rPDE relation for galaxies in three bins of $\Delta$SFR, representing galaxies that are either at least a factor of two below the SFMS (bottom panel), within a factor of two of the SFMS (middle panel) or at least a factor of two above the SFMS (top panel).  Looking first at the `normal' star-forming galaxies that lie within a factor of two of the SFMS (Figure \ref{pde_dsfr}, middle panel), the spaxels in these galaxies are well described by a linear slope (as determined from the ODR fit), in excellent agreement with the Ostriker \& Kim (2022) prediction.   We conclude that for typical star-forming galaxies in the extended ALMaQUEST sample the PRFM model agrees with the data, consistent with previous studies of normal star-forming galaxies at low redshift (e.g. Herrera-Camus et al. 2017; Sun et al. 2020a, 2023; Barrera-Ballesteros et al. 2021a).

However, significant deviations from the model are seen for ALMaQUEST galaxies that are either below (Figure \ref{pde_dsfr}, bottom panel) or above (Figure \ref{pde_dsfr}, top panel) the SFMS.  For galaxies located below the global SFMS (Figure \ref{pde_dsfr}, bottom panel), we see that the \sigsfr\ values lie below the PRFM model for all $P_{DE}$ (although the range is limited to values of log $P_{DE}/k_B < 5.0 $ K cm$^{-3}$).  In the context of the PRFM model, this suggests that there will be insufficient feedback from stellar processes to balance the pressure in the disk.  However, there are relatively few spaxels in the $\Delta$SFR$<-$0.3 dex bin and they hence contribute relatively little to the full sample show in the top left panel of Figure \ref{prfm_fig}.

On the other hand, over half of the 19,999 spaxels in the full star-forming sample are located in galaxies with $\Delta$SFR$>$+0.3 dex.  Half of the galaxies in this sample are central starbursts and one is a merger (i.e. as defined in Section 2), with several others exhibiting possible disturbances that were sufficiently ambiguous to not be classified as mergers.  However, the majority of the $\Delta$SFR$>$+0.3 dex sample do not show signs of interactions. Containing over 11,000 spaxels, the $\Delta$SFR$>$+0.3 dex sample therefore contributes significantly to the deviation between the data and the PRFM model in the top left panel of Figure \ref{prfm_fig}.

From the top panel of Figure \ref{pde_dsfr} we can see that the model and the data diverge in two distinct ways in different regimes.  First, at log $P_{DE}/k_B < 5.0 $ K cm$^{-3}$ the \sigsfr\ values are significantly \textit{higher} than the model would predict.  Conversely, at higher values of $P_{DE}$, the observed \sigsfr\ falls below the model prediction.  Combined, these effects lead to a gradient in the rPDE relation that is significantly more shallow in the high SFR sample than predicted by the model.  A hint of these low \sigsfr\ values at high $P_{DE}$ is even present in the `normal' star forming sample (Figure \ref{pde_dsfr}, middle panel), as well as in the EDGE-CALIFA sample (Barrera-Ballesteros et al. 2021a and our Figure \ref{lit_fig}), but it is only with the significant sample of highly star-forming galaxies in ALMaQUEST that the signal becomes very clear.

\begin{figure}
	\includegraphics[width=9cm]{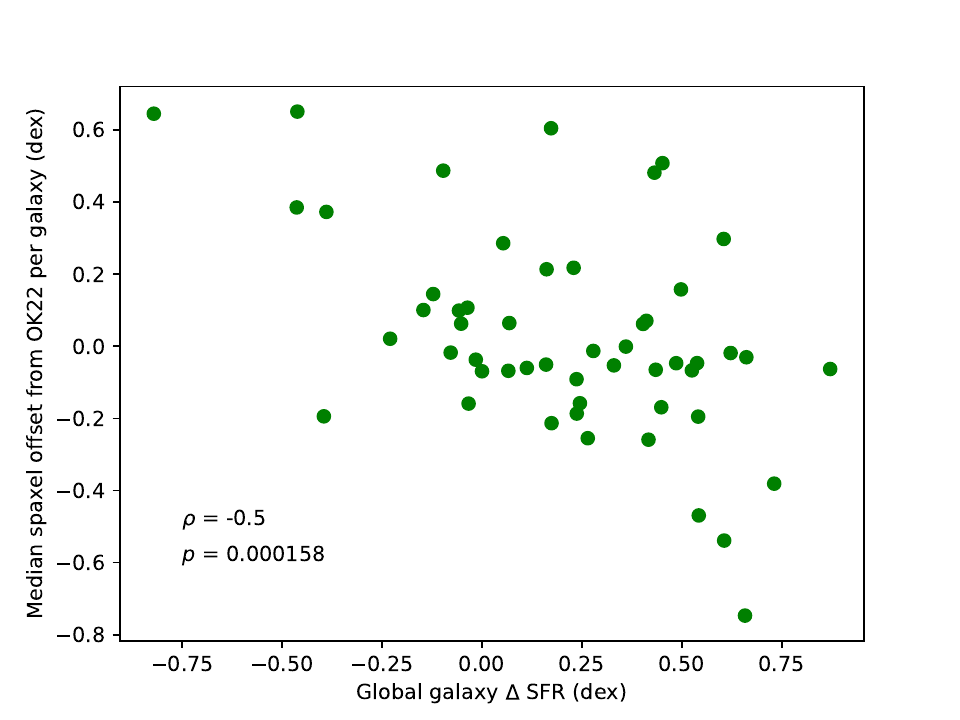}
        \caption{For all star-forming spaxels in a given galaxy, we compute the median offset of the measured $P_{DE}$ compared to the prediction of Ostriker \& Kim (2022) and plot the result as a function of $\Delta$ SFR.  The anti-correlation (Pearson correlation test $p$ and $\rho$ values reported in the lower left of the figure) supports the results in Figure \ref{pde_dsfr} that whilst the pressure regulated feedback modulated formalism is a good description of normal star-forming galaxies, it breaks down for galaxies above and below the SFMS. }
    \label{dpde_dsfr}
\end{figure}

To further dissect the dependence of the rPDE relation on $\Delta$SFR in finer detail than the three bins shown in Figure \ref{pde_dsfr}, for each galaxy we compute the median spaxel offset from the Ostriker \& Kim (2022) relation.  Figure \ref{dpde_dsfr} shows that there is indeed an anti-correlation between the offset between the data and the PRFM prediction and $\Delta$ SFR (Pearson correlation test results are given in the lower left of the panel), indicating that whilst the pressure regulated feedback modulated formalism is a good approximation for main sequence galaxies, it breaks down for galaxies with more extreme star formation rates.  Figure \ref{dpde_dsfr} also therefore indirectly shows that a single rPDE relation is not representative of the entire galaxy sample.

\begin{figure*}
	\includegraphics[width=19cm]{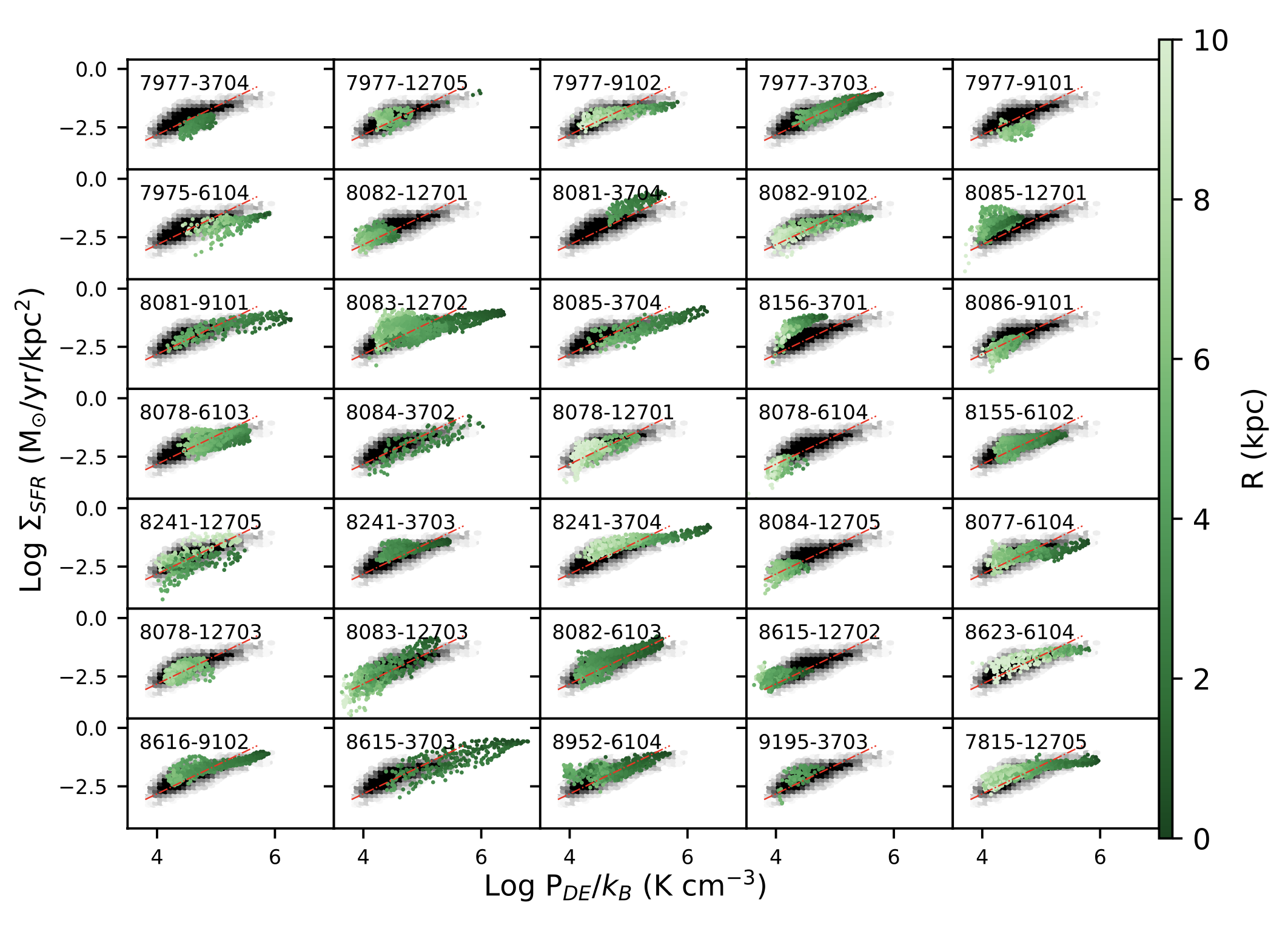}
        \caption{The rPDE relation for a random selection of 30 galaxies in the extended ALMaQUEST sample.   The MaNGA plate-IFU of each galaxy is noted in the top left corner.  The grey scale background shows the full sample of $\sim$20,000 star forming spaxels as a visual reference.  The red dot-dashed line shows the prediction from the TIGRESS simulation of Ostriker \& Kim (2022).}
    \label{prfm_pifu}
\end{figure*}

The non-universality of the rPDE relation can be seen most effectively by examining individual galaxies.  In Figure \ref{prfm_pifu} we show a selection of galaxies from the extended ALMaQUEST sample; these are chosen randomly, but with the (arbitrary, for visual presentation purposes) requirement that the galaxy have at least 100 star-forming spaxels.  Although this minimum spaxel count means that green valley galaxies are under-represented in this montage, it does not affect the qualitative point that different galaxies exhibit different rPDE relations.  In each panel we also show the theoretical relation from Ostriker \& Kim (2022) as a red dot-dashed line and the ensemble of all star-forming spaxels, reproduced from the top left panel of Figure \ref{prfm_fig} as background greyscale for reference.  As has been previously shown for the other three star formation scaling relations (rSK, rMGMS and rSFMS;  Ellison et al. 2021a; Pessa et al. 2021, 2022; Casasola et al. 2022) the rPDE for the galaxies in ALMaQUEST can lie significantly above or below both the ensemble and the linear theoretical relation.  Considerable differences in slope are also observed, although the flattening at high values of $P_{DE}$ is a common feature.  We conclude that the rPDE relation is not universal between galaxies.  Nonetheless, it may still be the case that there is less variation in the rPDE relation than for other star formation scaling relations (rSK, rSFMS and rMGMS).  We turn to this point in the next sub-section.

\begin{figure}
	\includegraphics[width=9cm]{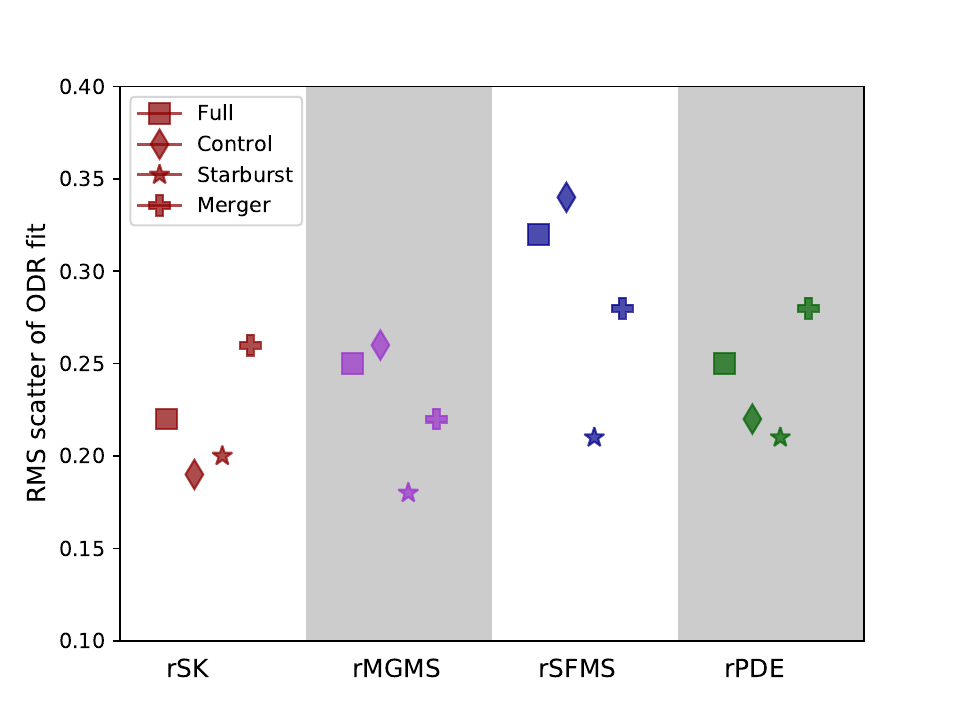}
        \caption{The RMS scatter around the ODR fit derived for each of four different sub-samples (distinguished by symbols) and four different scaling relations (distinguished by colours).  All of the relations have scatters within $\sim$ 0.2 -- 0.3 dex (depending on the galaxy sub-sample), with the rPDE exhibiting no tighter scatter than the others. }
    \label{rms_fig}
\end{figure}

\begin{figure}
	\includegraphics[width=9cm]{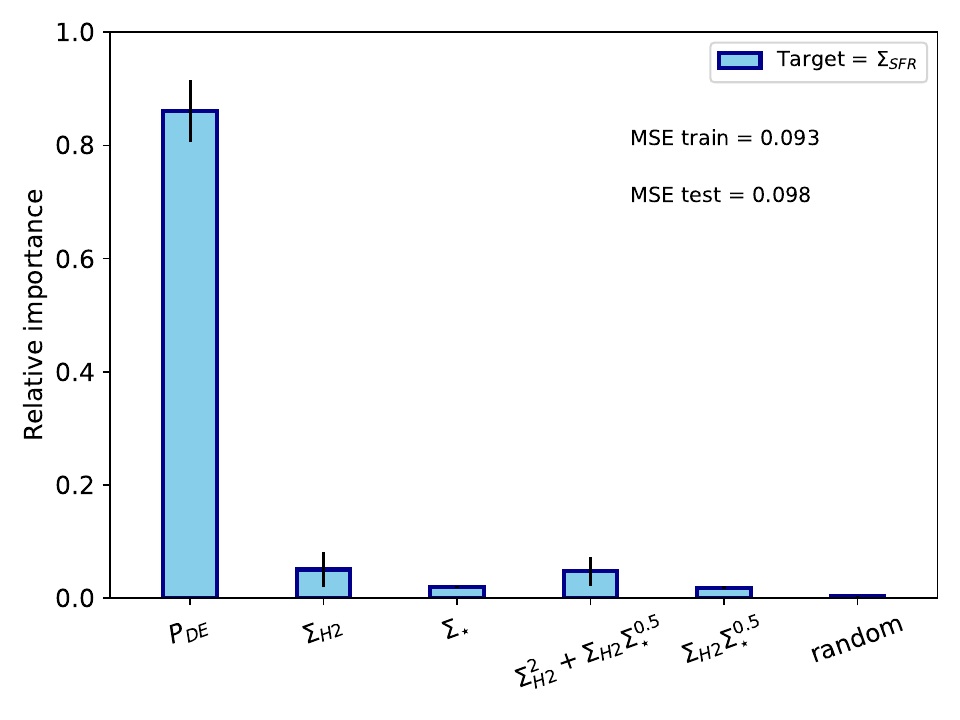}
        \caption{The relative importance of six variables in predicting \sigsfr\ as determined from a random forest regression analysis.  $P_{DE}$ is a far superior predictor of \sigsfr\ than either \sigstar\ or \sigh2, and similarly better than other combinations of these variables. }
    \label{rf_fig}
\end{figure}

\subsection{Is the rPDE relation the most fundamental star formation scaling relation?}

Scaling relations are often used as a way to motivate the understanding of physical processes.  However, we are well-drilled in the caveat that correlation does not imply causation, stimulating considerable effort in disentangling fundamental correlations from those that arise as by-products.  This has been achieved with a variety of statistical methods, including classical analyses such as the assessment of scatter and correlation, dependence on additional variables and partial correlation coefficients, to more sophisticated approaches that attempt to rank the relative importance of variables to a target (e.g. Ellison et al. 2008b; Teimoorinia et al. 2016; Dey et al. 2019; Bluck et al. 2020, 2022; Ellison et al. 2020a,b; Baker \& Maiolino 2023).

Specifically, there has been active discussion in the literature concerning the star formation scaling relations studied in this paper.   Lin et al. (2019) were the first to suggest that the rSFMS is not fundamental, but rather a result of combining the rMGMS and rSK.    Lin et al. (2019) propose that the rMGMS and the rSK relation set-up a 3-dimensional correlation that projects onto the plane of \sigstar -- \sigsfr\ to produce the rSFMS, despite there being no direct causal connection between these two variables.  Several other authors have since supported these conclusions, by replicating the 3-dimensional structure in other datasets (e.g. S\'{a}nchez et al. 2021) finding larger scatter in the rSFMS compared with the other relations (Ellison et al. 2021a; Morselli et al. 2020; Pessa et al. 2021) and no statistical evidence for a relation between \sigstar\ and \sigsfr\ once the rMGMS and rSK are accounted for (Baker et al. 2022).

The PRFM theory moves beyond these simple two-variable scaling relations by predicting that the ISM pressure captures the fundamental regulator of star formation rate.  This conclusion was supported by Barrera-Ballesteros et al. (2021a) in their study of EDGE-CALIFA data and they conclude  `Our results also suggest that hydrostatic pressure is the main parameter that modulates star formation at kpc scales, rather than individual components of the baryonic mass.'  In order to test this conclusion with the extended ALMaQUEST data, in Figure \ref{rms_fig} we plot the RMS scatter of the data around the ODR fit for the four different scaling relations (shown in distinct colours in vertical groupings) and in four different galaxy sub-samples (shown as different symbols).  We find that the rPDE relation shows no less scatter (around the linear fit) than the other three relations.  Moreover, as expected based on the considerable galaxy to galaxy variation seen in Figure \ref{prfm_pifu}, the scatter is variable depending on the sample, because each one (even the control sample) has a range of star formation properties.

Although a comparison of scatters has been previously used to assess the `fundamental' nature of scaling relations (e.g. Lin et al. 2019; Morselli et al. 2020; Ellison et al. 2021a; Pessa et al. 2021), the relationship between $P_{DE}$ and \sigsfr\ is clearly not linear.  This non-linearity, which manifests as a flattening of the relation at high pressures (which tend to be associated with highly star-forming galaxies) was not obvious in previous studies which were limited to main sequence galaxies (e.g. Herrera-Camus et al. 2017; Sun et al. 2020a; Barrera-Ballesteros et al. 2021a).  Using the scatter around a linear fit is therefore unlikely to be the optimal way to assess the fundamental nature of the rPDE relation, motivating a more sophisticated treatment of the data.

We therefore turn to a random forest analysis, which is powerful at extracting non-linear dependences in data.  A random forest consists of a series of decision trees which can be used to assess the relative importance of input variables in determining a given target variable.  Our approach follows closely the analyses presented in our previous works (e.g. Baker et al. 2022, 2023); in particular, we refer the curious reader to Appendix B of Bluck et al. (2022) for a detailed explanation of the random forest methodology and mathematical formulation.  

The target variable of our random forest is \sigsfr.  We first assess the parameters that represent the scaling relations examined in this paper, namely \sigh2\ (i.e. testing the rKS relation), \sigstar\ (i.e. testing the rSFMS) and $P_{DE}$ (i.e. testing the rPDE relation).  This random forest is therefore an extension of the work presented by Baker et al. (2022) who assessed only the relative importances of \sigh2\ and \sigstar, finding that the former was more important for predicting \sigsfr.  The relative performance of  $P_{DE}$, \sigh2\ and \sigstar\ and for predicting \sigsfr\ is shown in the first three bars of Figure \ref{rf_fig}, where the error bars represent the variance in 100 independent realizations of the training, validation and testing data.  The mean squared error (MSE) is reported in the top right of the figure for both the training and testing data.  From the first three bars of Figure \ref{rf_fig} it is clear that the $P_{DE}$ is much more important than either \sigh2\ or \sigstar\ for determining \sigsfr, and that this result is highly significant and stable. We emphasize that the results presented in Figure \ref{rf_fig} only rank the relative importances of the features included in our test set, and by design, add up to a total of one.  Therefore, these relative importances are not absolute values, and their quantitative values are only relevent for the fixed feature set included in our test.  For example, the ratio of the relative importances of any two variables shown in Figure \ref{rf_fig} would change if more (or less, or different) features were included.

In order to assess whether the successful performance of $P_{DE}$ as a predictor of \sigsfr\ is a `trivial' result, in the sense that $P_{DE}$ is itself a combination of \sigh2\ and \sigstar, we include in the random forest two additional variables that are also combinations of \sigstar\ and \sigh2.  The first is a modified version of $P_{DE}$, in that it combines \sigh2\ and \sigstar\ in the same relative proportions as $P_{DE}$ (Equation \ref{prfm_eqn}), but without the addition of \sighi\ (which we recall we have set to be a constant), velocity dispersion nor information on $R_{50}$ and without any coefficients (such as physical constants).  That is, we compute  \sigh2$^2$  + \sigh2 \sigstar$^{0.5}$.  The other extra variable we include is also a combination of \sigstar\ and \sigh2\ that represents the `extended' Kennicutt-Schmidt relation of Shi et al. (2011).  These authors showed that \sigh2\ \sigstar$^{0.5}$ correlated more tightly with \sigsfr\ than \sigh2\ alone.  These two additional variables are shown in the 4th and 5th bars in Figure \ref{rf_fig}; for reference, a random variable is shown in the final bar.  All three of these additional variables have a very small relative importance compared with $P_{DE}$.

Taken together, the results in Figure \ref{rf_fig} show that $P_{DE}$ represents a physically meaningful combination of \sigstar\ and \sigh2\ for the prediction of \sigsfr.  Neither the extended KS law, nor a modification of the $P_{DE}$ equation (which also combines \sigstar\ and \sigh2) perform anywhere near as well as $P_{DE}$. Indeed, after P$_{DE}$ it is the single variable \sigh2\ that has the next highest relative importance (but it is in a very distant second place).   These results reflect the ability of the random forest analysis to extract non-linear relations.  \textit{Our random forest therefore demonstrates that $P_{DE}$ is a far superior predictor of \sigsfr\ than \sigh2\ or \sigstar\ alone}, a result that was not seen when assessing the scatter around the linear relations alone (Figure \ref{rms_fig}).

It is also important to recognize that, whilst there are a number of uncertainties associated with our calculation of $P_{DE}$ (which we explore in detail in the next section) any improvement in the accuracy of the measurement of the pressure will only increase the dominance shown in Figure \ref{rf_fig}.  Put another way, $P_{DE}$ is the most predictive parameter (of those tested) for \sigsfr, despite some uncertainties in its derivation.  As a final comment, we note that our analysis says nothing about whether $P_{DE}$ is an \textit{optimal} combination of \sigstar\ and \sigh2\ (or any other additional variables), only that it is better than the individual variables (or the other combinations) we have tested.

\subsection{Uncertainties in the measurement of $P_{DE}$}

We have so far shown that dynamical equilibrium pressure is the more predictive of the star formation rate surface density than either \sigh2\ or \sigstar\ (Figure \ref{rf_fig}), but that the relationship between $P_{DE}$ and \sigsfr\ shows significant deviations from the prediction of the PRFM model (Figures \ref{prfm_fig} -- \ref{prfm_pifu}) .  In this sub-section, we consider whether the assumptions used in our calculation of $P_{DE}$ might explain the observed discrepancy between the PRFM model and the extended ALMaQUEST data, with a particular focus on the highly star-forming galaxies in the sample shown in the top panel of Figure \ref{pde_dsfr}.  We remind the reader that the nature of the disagreement is that the slope of the rPDE relation is much flatter in the data than predicted by the model, such that \sigsfr\ is too high for its $P_{DE}$ at values of log $P_{DE}/k_B < 5$ K cm$^{-3}$ and \sigsfr\ is lower than the model at higher pressures.

\subsubsection{Assumption of a fixed \sighi}

First, we consider our assumption of a constant value of \sighi\ = 7 M$_{\odot}$ pc$^{-2}$ for all spaxels.   At high values of $P_{DE}$ we do not expect our choice of \sighi\ to be a significant issue.  In order to bring the data into agreement with the model in the high $P_{DE}$ regime, the pressure needs to be reduced from its current value.  This requires the atomic surface density to be lower than our currently assumed value of \sighi\ = 7 M$_{\odot}$ pc$^{-2}$.  However, as noted earlier, the typical H$_2$ surface densities for ALMaQUEST spaxels are at least \sigh2\ = 10 M$_{\odot}$ pc$^{-2}$, meaning that our assumed value of \sighi\ does not significantly contribute to $\Sigma_{gas}$.  Indeed, in the extended ALMaQUEST sample, $\Sigma_{gas}$ will be dominated by \sigh2\ for the majority of our spaxels for any reasonable choice of \sighi\ below the saturation threshold (\sighi\ $\sim$ 9 M$_{\odot}$ pc$^{-2}$) observed in nearby disks (Bigiel et al. 2008).  

However, at low $P_{DE}$ values, where the pressure needs to be increased in order to bring it into agreement with the data, it is reasonable to contemplate whether our assumed value of \sighi\ might be too low.  We experimented with changes in the adopted value of \sighi, as well as including a randomized element to emulate a range of HI surface densities (e.g. following Barrera-Ballesteros et al. 2021a).  For values of \sighi\ up to the locally observed saturation threshold, there is little impact and the data remain inconsistent with the model.  We have to increase the atomic contribution to \sighi\ $\sim$ 14 M$_{\odot}$ pc$^{-2}$ before there is some reasonable agreement between the data and the model at low $P_{DE}$.  However,  such high values of \sighi\ simultaneously exacerbate the discrepancy between the model and the data at high pressures, and so are not a good solution for the data as a whole.  Moreover, there is little evidence to empirically (or theoretically) motivate such a high value of \sighi\ for our sample.  Despite the more intense UV radiation fields that come with high SFRs (that might be expected to dissociate H$_2$), this is offset by higher gas volume and surface densities, which shield against far ultra-violent radiation and promote H$_2$ formation. Consequently, models and observations alike find that \sighi\ is predominantly sensitive to metallicity, rather than SFR, with \sighi\ $>$10 M$_{\odot}$ pc$^{-2}$ only expected in regions of much lower metallicity than in our sample (Krumholz, McKee \& Tumlinson 2009; Fumagalli, Krumholz \& Hunt 2010; Schruba, Bialy \& Sternberg 2018).  Indeed, Bigiel et al. (2008) find that, even in the centres of galaxies with locally enhanced \sigsfr, the \sighi\ is rarely in excess of $\sim$ 8 M$_{\odot}$ pc$^{-2}$.

As a final comment on the choice of a fixed \sighi, we note that Barrera-Ballesteros et al. (2021a) made a similar assumption for the EDGE-CALIFA sample, although those authors additionally added a random component to the value of \sighi\ = 7 M$_{\odot}$ pc$^{-2}$ to emulate the range of observed values.  Despite this similar approach, Barrera-Ballesteros et al. (2021a) find that their data \textit{do} agree with the model at low pressures, in much the same way that the `normal' star-forming galaxies in ALMaQUEST do (middle panel of Figure \ref{pde_dsfr}).  Overall, we conclude that the choice of a fixed \sighi\ is unlikely to be the reason that our data disagree with the PRFM model.

\subsubsection{Assumption of a fixed CO-to-H$_2$ conversion factor}

Next, we consider the impact of using a fixed CO conversion factor.  This is almost certainly an incorrect assumption, as $\alpha_{CO}$ is known to vary as a function ISM conditions, including density, galactocentric radius and metallicity (Narayanan et al. 2012; Bolatto 2013; Sandstrom et al. 2013; Accurso et al. 2017; Hunt et al. 2020; Gong et al. 2020). However, before embarking on an exploration of alternative values of $\alpha_{CO}$, we begin by considering the general impact that this alteration would have on our results (beyond the change to $P_{DE}$).  Adopting a different (or variable) $\alpha_{CO}$ has the obvious direct result of changing \sigh2, which we (and others) have shown has a strong and tight correlation with \sigstar\ (e.g. Fig \ref{mgms_fig}).  Changing the conversion factor will therefore have a direct impact on our observed rMGMS.  As shown in Fig \ref{mgms_fig}, unlike the rPDE relation, the rMGMS of the extended ALMaQUEST is very similar for central starburst and control galaxies.  Moreover, the slope of the rMGMS is identical for galaxies with $\Delta$ SFR$>0.3$ dex and those with $-0.3< \Delta$ SFR $<0.3$ dex, and yet the rPDE relation is very different in these two regimes (Fig. \ref{pde_dsfr}).  Although there is a small offset to higher \sigh2\ at fixed \sigstar\ (i.e. higher gas fractions) in the $\Delta$ SFR$>0.3$ dex sample by 0.1 dex, this offset is modest and higher gas fractions are in any case frequently observed for galaxies above the main sequence (e.g. Saintonge et al. 2012, 2016).  Any significant reduction in $\alpha_{CO}$ (as might be more appropriate for a starburst sample, and necessary to bring the high $P_{DE}$ values into line with the Ostriker \& Kim 2022 relation) would lead to starburst galaxies with suppressed gas fractions, which is not consistent with either other observations or the fuel requirements for their high SFRs.  In short, the rMGMS gives us a prior hint that significant reductions to $\alpha_{CO}$ are likely inappropriate for our sample.

Nonetheless, we next embark on a quantitative assessment of alternative values of $\alpha_{CO}$.  First, we re-compute $P_{DE}$ using the metallicity dependent conversion factor given in Equation \ref{alpha_eqn}.  Since the majority of the spaxels in our sample have 8.55 $<$ 12 + log O/H $<$ 8.75, adopting Equation \ref{alpha_eqn} leads to conversion factors slightly higher than our fiducial value of 4.3.  As a result, the recomputed values of \sigh2\ (and hence $P_{DE}$) are slightly larger than those derived from our original values with fixed $\alpha_{CO}$, bringing the spaxels with low $P_{DE}$ into marginally better agreement with the model, but the change is small and the disagreement with the model persists.  In the regime of high $P_{DE}$, where the pressure is too large compared with the model, the metallicity dependent conversion factor exacerbates the discrepancy between the model and the data.  We also test the Accurso et al. (2017) conversion prescription which, in addition to a primary dependence on metallicity, has a mild secondary dependence on galaxy offset from the main sequence.  The addition of this second term has a minimal effect on our data.

As an alternative to the metallicity dependent conversion factor, we next consider whether choosing a lower $\alpha_{CO}$ (than our fiducial value of 4.3) could resolve the discrepancy.  Values as low as $\alpha_{CO} \sim 1$ (or even smaller) are seen in some parts of disks as well as in extreme starbursts (e.g. Bolatto et al. 2013; Sandstrom et al. 2013; Teng et al. 2023).   Artificially lowering the conversion factor by only 50\% is sufficient to bring the high $P_{DE}$ spaxels into line with the model; but the disagreement at low $P_{DE}$ is then exacerbated.  However, it is likely that different values of $\alpha_{CO}$ are applicable in different galactic regions, such that a blanket adjustment of $\alpha_{CO}$ is also probably not correct (even for starbursts).  There are several formulations for a variable $\alpha_{CO}$ in the literature that we can potentially adopt (e.g. Narayanan et al. 2012; Bolatto et al. 2013; Gong et al. 2020; Teng et al. 2023). We test the implementation of two of these, namely a dependence of $\alpha_{CO}$ on the total surface density $\Sigma_{tot}$ = \sigstar\ + \sigh2\ + \sighi\ with index of $-0.5$ (green line in Fig. 12 of Bolatto et al. 2013) and a dependence on the CO line width (Equation 6 from Teng et al. 2023).  Both of these prescriptions result in values of lower $P_{DE}$, but to the extent that the starburst data now lie a factor of 2--3 to the left of the PRFM model (essentially `over-correcting' the original discrepancy seen between the model and the data).

Finally, we assess the impact of simply removing spaxels with physical properties that might be indicative of more extreme conversion factors.  We try removing spaxels with line widths larger than 20 \kms\ (as Teng et al. 2023 show that $\alpha_{CO}$ decreases when the line is broad) as measured from the moment-2 maps, removing spaxels where $\Sigma_{tot} >$ 300 M$_{\odot}$ pc$^{-2}$ (Bolatto et al. 2013's  $\alpha_{CO}$ prescription scales as $\Sigma_{tot}^{-0.5}$) and removing spaxels within 2 kpc of the galaxy centre (Sandstrom et al. 2013 and Teng et al. 2023 identify central regions as most deviant from Galactic conversion factors).  The resulting rPDE relation for starburst galaxies after each of these three purges still shows the same general trend (albeit with fewer spaxels) as shown in the top panel of Figure \ref{pde_dsfr}.  We conclude that whilst alternative values of $\alpha_{CO}$ could be hand-picked to bring any given spaxel in our sample into alignment with the PRFM model, we have not identified a physically motivated formalism that can achieve this. 

All of the factors considered thus far in our discussion have focussed on elements that contribute to the calculation of \siggas, i.e. uncertainties in either \sighi\ or \sigh2.  Although we have recognized some limitations in our ability to compute \siggas\ to high accuracy, we have also not found evidence that the uncertainties in the gas contribution is driving the disagreement between the PRFM model and the data.  We also note that in the calculation of $P_{DE}$ (i.e. Equation \ref{prfm_eqn}) the first term (the contribution to the pressure from gas self-gravity) is almost always smaller (by a factor of two, on average) than the second (stellar) term.  Therefore, modest changes to either \sighi\ or \sigh2\ have a minor impact on the calculation of  $P_{DE}$.

\subsubsection{Assumption of a fixed gas velocity dispersion}

The next uncertainty we consider is our assumption of a fixed $\sigma_{gas,z}$ = 11 \kms.  Although we can not measure this parameter accurately in our own data, the assumed value is consistent with the fairly universal value for normal star-forming disks at low redshift (e.g.  Kennicutt \& Evans 2012; Caldu-Primo et al. 2013).  Thanks to the availability of higher resolution data, Sun et al. (2020a) were able to explicitly assess the impact of this choice on kpc-scale data, finding that it could lead to a small over-estimate in $P_{DE}$, but with a scatter that was generally within 0.2 dex.  Although a downward correction by 0.2 dex in $P_{DE}$ somewhat relieves the tension between the data and the model at high pressures, the $P_{DE}$ would need to be reduced by values closer to 0.5 -- 1 dex in order to be fully in line with the theory.  Moreover, a downward correction in $P_{DE}$ acts in the contrary direction needed to resolve the discrepancy between the data and the model in the low $P_{DE}$ regime.  In order to match the data to the theory in the low pressure regime, the calculated pressure needs to increase, a result which might be achieved if our assumed value of  $\sigma_{gas,z}$ = 11 \kms\ is an under-estimate.  Since elevated velocity dispersions are indeed expected for galaxies with high SFRs (Krumholz \& Burkhart 2016; Krumholz et al. 2018), we re-compute $P_{DE}$ using the actual (line of sight) values of the velocity dispersions as measured from the CO(1-0) line.  Although the median value in the data is $\sim$ 11 \kms\ (which is also the spectral resolution of the ALMaQUEST data), there is a tail to values as large as 80 \kms.  Of course, since we have not modelled and removed disk rotation, these velocity dispersions represent an upper limit to the possible value of $\sigma_{gas,z}$, but this nonetheless serves our purpose of assessing whether our assumption of $\sigma_{gas,z}$ = 11 \kms\ is causing the discrepany between the data and the model at low pressures.  We find that using the measured values of the CO velocity dispersion does not reduce the offset between the model and theory in high SFR galaxies.

\subsubsection{Uncertainties in \sigsfr}

Finally, we can also consider whether the discrepancy between the PRFM model and the ALMaQUEST data might be due to incorrect values of  \sigsfr.   At high values of $P_{DE}$ the discrepancy between the model and the data could be explained if \sigsfr\ has been under-estimated, for example due to high extinction.  Indeed, it has been suggested that the Balmer decrement method (although very widely used) might significantly under-estimate extinction (Inoue et al. 2001; Dopita et al. 2003).  However, in general, the method we have adopted to determine \sigsfr, which is based on dust-corrected H$\alpha$ emission, has been found to be a good match to those determined from the UV and IR in the CALIFA DR2 (Garcia-Benito et al. (2015) which includes $\sim$ 10 percent mergers in various stages (Catalan-Torrecilla et al. 2015).      A visual inspection of the high SFR galaxies reveals that they are not obviously dusty or highly inclined.  Moreover, high values of \sigsfr\ do exist in our sample for some galaxies (e.g. Ellison et al. 2021a), demonstrating that we can identify such regions and note that the distinctive flattening seen in the rPDE relation (top left panel of Figure \ref{prfm_fig}) is not seen in the rSK relation (top left panel of Figure \ref{ks_fig}), indicating that the raw variables of \sigsfr\ and \sigh2\ are well-behaved.  In terms of a possible over-estimate of \sigsfr\ that might cause disagreement with the PRFM model at low values of $P_{DE}$ we re-iterate that we have removed spaxels with contributions from AGN.  Therefore, the measured H$\alpha$ flux should not have significant contributions from non-stellar sources, indicating that an over-estimate of \sigsfr\ is unlikely to be the cause of the disagreement between the model and data in the low $P_{DE}$ regime. 

\medskip

In summary, we have not identified any methodological cause for the disagreement between the observed rPDE relation in the extended ALMaQUEST sample and the PRFM model.  It appears that, whereas the PRFM model is a good representation of the data for galaxies on the SFMS (middle panel of Figure \ref{pde_dsfr}), there is a persistent tension between the extended ALMaQUEST sample and the theory for high SFR galaxies (top panel of Figure \ref{pde_dsfr}).  And yet we have also found that $P_{DE}$ is the most (of the variables tested) relevant predictor of star formation in our sample.  Taken together, these results indicate that whilst dynamical equilibrium pressure is a key variable in regulating star formation, the PRFM model is an incomplete formulation of the relation.  In the final sub-section below, we consider alternative models that might better represent the full diversity present in the extended ALMaQUEST sample.

\begin{table*}
  \caption{Symbol definitions used in the Krumholz et al. (2018) star formation model (Equations \ref{eqn30} -- \ref{eqn20}) and their adopted values, where relevant.}
\begin{tabular}{lcl}
\hline
Symbol & value & Meaning \\
\hline                                                                                                      
$\Omega$ & ... & Galaxy angular velocity \\
$v_{\phi}$ & ... & Galaxy rotation curve velocity \\
$t_{orb}$ & ... & Galaxy orbital period \\
$f_{sf}$ & ... & Fraction of ISM in molecular phase\\
$\rho_{min}$ & ... & Minimum midplane density required to produce rotation curve\\
$\beta$ & 0 & Rotation curve index\\
$f_{g,Q}$ & 0.5 & Fractional contribution of gas to Toomre $Q$\\
$f_{g,P}$ & 0.33 & Fractional contribution of gas to mid-plane pressure\\
$\phi_{mp}$ & 1.4 & Ratio of total pressure to turbulent pressure at midplane\\
$\epsilon_{ff}$ & 0.015 & Star formation efficiency per free-fall time\\
$t_{sf,max}$ & 2 Gyr & Maximum star formation timescale\\

\hline
\end{tabular}
\label{k18_tab}
\end{table*}

\subsection{What causes the discrepancy between the PRFM model and the ALMaQUEST data?}

We have found that the extended ALMaQUEST data deviate from the PRFM model most dramatically for spaxels in highly star-forming galaxies.   Hassan, Ostriker \& Kim (in prep) have recently presented a new analytic expression for $P_{DE}$ specifically designed to account for ISM conditions in starburst galaxies and/or those with particularly turbulent ISM conditions (e.g. Fisher et al. 2019).  In particular, the new formulation guards against an over-prediction of the stellar (second) term in Equation \ref{prfm_eqn} if the velocity dispersion is very large and/or the gas scale height exceeds that of the stars.  We have re-computed $P_{DE}$ using the new formulation of Hassan et al. (in prep) and find that $P_{DE}$ is typically reduced by $\sim$ 5\% compared with the fiducial calculation using Equation \ref{prfm_eqn}.  The disagreement between our highly star-forming galaxies and the PRFM model is therefore not resolved by using this new formalism.

Having explored a number of possible avenues for reconciliation between the ALMaQUEST data and the PRFM theory, we have found no obvious culprit amongst observational assumptions, nor in the most recent version of the PRFM expression that should be most suitable for our high SFR galaxies.   We are therefore left with the conclusion that, although the PRFM model is a good representation of the data for galaxies that lie on the SFMS (see also Barrera-Ballesteros et al. 2021a; Sun et al. 2020a, 2023), the model exhibits a fundamental disagreement with galaxies in the high (and low) SFR regime.  In the remainder of this section we consider why this might occur.

The PRFM model is predicated on the tenet that feedback from star formation is the sole source of turbulence in the ISM, and that this in turn balances the pressure across the galactic disk.  Starting with Ostriker \& Shetty (2011), the generations of models that have built on this assumption all ignore (by design) the larger scale physics of the galaxy.  If there are additional sources of turbulence, the underlying assumptions of the PRFM model are undermined.  Large scale motions of gas within galaxies could readily provide such additional turbulence, both through on-going gas accretion as well as radial flows, and there is an abundance of evidence that both of these effects must be at play in real galaxies.  For example, the well-established observation that galaxies have depletion times $\sim$ 1-2 Gyr (Leroy et al. 2008, 2013; Bigiel et al. 2008, 2011; Saintonge et al. 2011), i.e. much shorter than the Hubble time, and yet continue to form stars over an extended period demonstrates that on-going gas accretion/disk fuelling must be occuring.  A similar conclusion is reached from the observation that the HI gas mass density in damped Lyman $\alpha$ systems (gas-rich galaxies seen in absorption in quasar spectra) is flat over a large range in redshift, at least $1 <  z < 4$ (e.g. Zafar et al. 2013; S\'{a}nchez-Ramirez et al. 2016), despite high rates of cosmic star formation in this period.  Inflows of gas are responsible for central starbursts in both models (Barnes \& Hernquist 1991; Blumenthal \& Barnes 2018; Moreno et al. 2015, 2021) and observations (Ellison et al. 2008a, 2013; Scudder et al. 2012; Thorp et al. 2019; Bickley et al. 2021; Garay-Solis et al. 2023) of galaxy mergers, and starbursts in general (not just those that are merger driven) have preferentially enhanced SFRs in their centres (Ellison et al. 2018; Medling et al. 2018; Wang et al. 2019).  Multiphase outflows are also an apparently ubiquitous feature of low $z$ galaxies, driven by both AGN and star formation (Fluetsch et al. 2019; Roberts-Borsani \& Saintonge 2019; Avery et al. 2021).  Krumholz et al. (2018) have argued that these large scale effects can dominate over star formation as the primary source of turbulence.  Indeed, it is in the high SFR regime, where we see the strongest disagreement between the data and the PRFM model, that we might expect that these large scale effects to dominate.  Likewise, in galaxy mergers where there are not only radial gas flows but galaxy-wide disruptions, we see that the data are in relatively poor agreement with the PRFM model (lower right panel in Figure \ref{prfm_fig}).

Although Krumholz et al. (2018) do not make explicit predictions for the rPDE relation, it is nonetheless possible to extract the expected relationship from their formalism.  Our starting point is Equation 30 of Krumholz et al. (2018) who express the star formation rate as

\begin{equation}\label{eqn30}
  \Sigma_{SFR} = f_{sf} \Sigma_{gas} \frac{\epsilon_{ff}} {t_{ff}}
  \end{equation}

\noindent where $f_{sf}$ is the fraction of the total gas (\siggas) that is in the star-forming (molecular) phase and $t_{ff}$ and $\epsilon_{ff}$ are the free-fall time and star formation rate per free-fall time in this gas.  $\epsilon_{ff}$ is set to be 0.015, consistent with a wide range of environments and the best fit to the data derived by Krumholz, Dekel \& McKee (2012).  The ratio of free-fall time and SFR per free-fall time is parametrized in Krumholz et al. (2018) as the Toomre timescale (their Equation 31):

\begin{equation}\label{eqn31}
t_{sf,T} = \frac{t_{ff}}{\epsilon_{ff}} = \frac{\pi Q}{4 f_{g,Q} \epsilon_{ff}} \sqrt { \frac{3 f_{g,P} \phi_{mp}} {2(1+\beta)} } \frac{1}{\Omega}
\end{equation}

\noindent where the description of the variables and their adopted values are given in Table \ref{k18_tab}.  For the majority of these variables we have adopted the default values from Krumholz et al. (2018) with the exception that we set the fractional contribution of gas to the pressure, $f_{g,P}$, to be 0.33 (instead of the fiducial value of 0.5 used in Krumholz et al. 2018) since this is the median value in the extended ALMaQUEST dataset.  We also use a fixed value of $\beta = 0$, i.e. a flat rotation curve.

Since the angular velocity at radius $r$ is given by $\Omega = v_{\phi}/r$ and the orbital period is $t_{orb} = 2\pi r / v_{\phi}$, substituting in Equation \ref{eqn31} therefore yields the requisite expression for $t_{sf,T}$ for different $t_{orb}$:

\begin{equation}
t_{sf,T} = \frac{\pi Q}{4 f_{g,Q} \epsilon_{ff}} \sqrt { \frac{3 f_{g,P} \phi_{mp}} {2(1+\beta)} } \frac{t_{orb}}{2 \pi}.
\end{equation}

\noindent In the Krumholz et al. (2018) model, if $t_{sf,T}$ is shorter than $t_{sf,max} =$ 2 Gyr then stars form in a continuous medium, otherwise the gas appears to break up into individual molecular clouds (Bigiel et al. 2008; Leroy et al. 2008, 2013).  In order to capture these two different `modes', the SFR in Equation \ref{eqn30} can thus be re-written as

\begin{equation}\label{eqn32}
  \Sigma_{SFR} = f_{sf} \Sigma_{gas} max (t_{sf,T}^{-1},t_{sf,max}^{-1}).
  \end{equation}

Next we must evaluate $f_{sf}$.  This is achieved using the theoretical model originally laid out by Krumholz, McKee \& Tumlinson (2009), with improvements as presented in McKee \& Krumholz (2010) and Krumholz (2013), referred to as the KMT+ model.   These models require input values of a clumping factor (following Krumholz et al. 2009, we adopt a value of 5) and $\rho_{min}$, the minimum mid-plane pressure required to produce the rotation curve.  $\rho_{min}$ is computed using equation 51 of Krumholz et al. (2018):

\begin{equation}\label{eqn51}
\rho_{min} = \frac{v_{\phi}^2} {4 \pi G r^2} (2 \beta + 1).
\end{equation}

Since $v_{\phi}/r = 2\pi / t_{orb}$, Equation \ref{eqn51} can be re-written as

\begin{equation}
\rho_{min} = \left( \frac{2 \pi} {t_{orb}} \right) ^2 \frac{(2 \beta + 1)} {4 \pi G}.
\end{equation}

Finally, we need to express $P_{DE}$ in terms of \siggas, which is Equation 20 in Krumholz et al. (2018):

\begin{equation}\label{eqn20}
P_{DE} = \frac{\pi G} {2 f_{g,P}} \Sigma_{gas}^2 .
\end{equation}

\noindent $f_{g,P}$ packages the contributions of the pressure from gas, stars and dark matter into a single term (see Equation 20 of Krumholz et al. 2018), such that Equation \ref{eqn20} is equivalent to the formalism used by Ostriker \& Kim (2022).

\begin{figure*}
	\includegraphics[width=17cm]{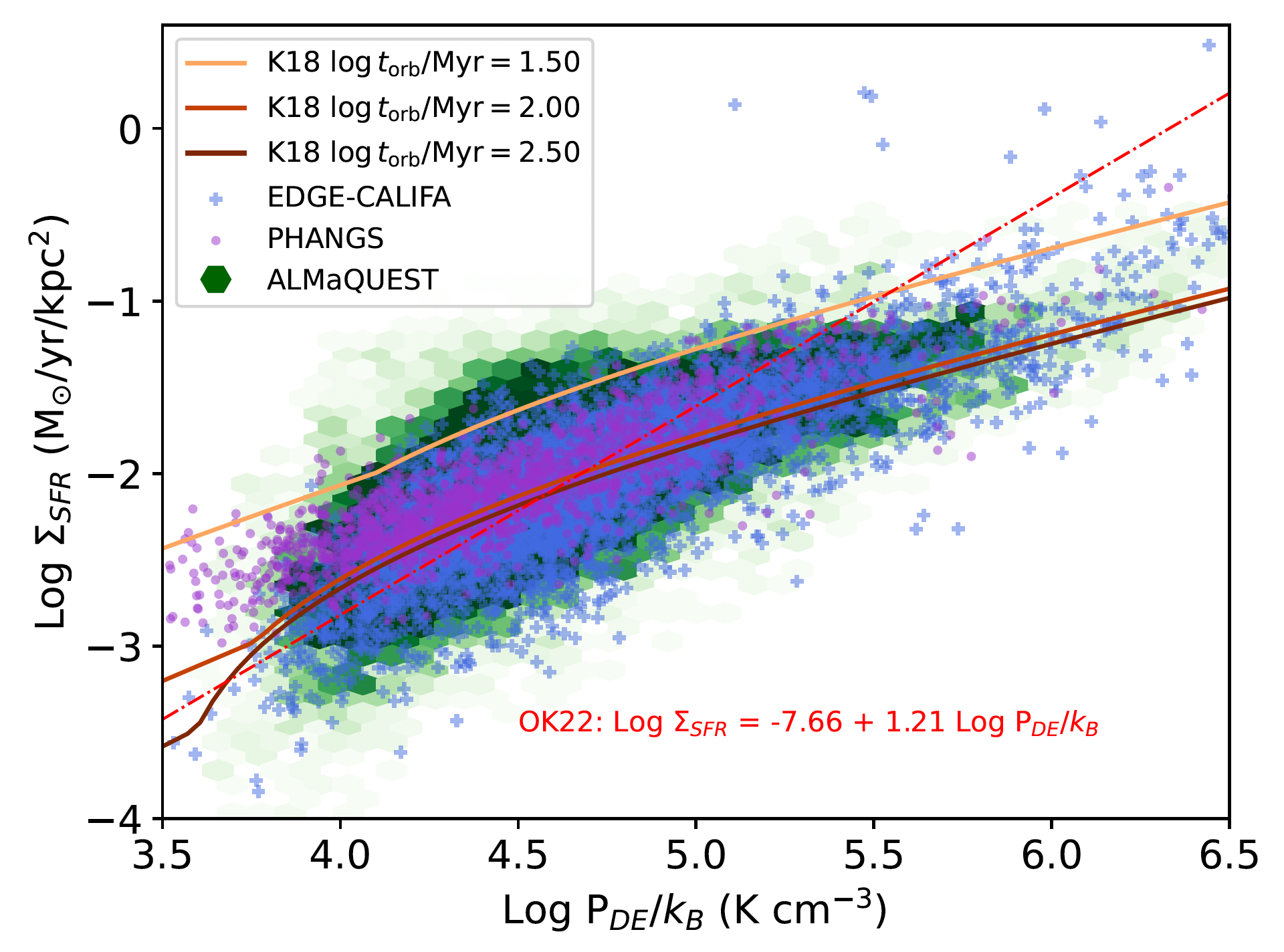}
        \caption{The rPDE relation for all star-forming spaxels in the extended ALMaQUEST dataset (the same data as shown in the top left panel of Figure \ref{prfm_fig}; green 2D histogram), data from the EDGE-CALIFA survey (Barrera-Ballesteros et al. 2021a; blue crosses) and PHANGS (Sun et al. 2020; purple points).  The Ostriker \& Kim (2022) PRFM theory prediction is shown with the red dot-dashed line; the data from all three datasets diverge from this model at high values of $P_{DE}$.  The orange curves show three examples of an alternative star formation model presented by Krumholz et al. (2018) model, for different choices of orbital times that are expected to encapsulate the range of conditions in the ALMaQUEST sample.  The flatter slope of the Krumholz et al. (2018) models is a better representation of the data at high $P_{DE}$ than the Ostriker \& Kim (2022) model.  Our results therefore support the idea that turbulence is injected in the ISM not only by feedback from star formation, but additionally through radial gas inflows that act to decrease the orbital time (Krumholz et al. 2018).  This effect is seen most strongly in mergers and starbursts, where the orbital times are shorter than in normal star-forming disks. }
    \label{k18_fig}
\end{figure*}

With these equations we can now evaluate \sigsfr\ in Equation \ref{eqn32} for different choices of the orbital time which essentially capture the relative contribution of feedback from non-stellar sources.  In Figure \ref{k18_fig} we show the Krumholz et al. (2018) model predictions for the rPDE relation for three representative values of $t_{orb}$, where the longer values are representative of normal disks and higher values more appropriate for mergers and central starbursts.  It can be seen that the flatter slope predicted by the Krumholz et al. (2018) model is in much better agreement with the data at high $P_{DE}$ than the PRFM model of Ostriker \& Kim (2022).  Indeed, reasonable choices of the orbital time result in predictions that encompass all of the data.  Adopting different choices for other variables modifies the curves shown in Figure \ref{k18_fig} slightly, but we find them to generally bracket the spread of the data.  We conclude that the extended ALMaQUEST dataset, in sampling more extreme galaxy environments, particularly in the high pressure regime, can provide important leverage for distinguishing star formation scenarios and favour a picture in which turbulence does not (always) come from star formation alone.

\section{Conclusions}\label{conclusions_sec}

We have presented the extended ALMaQUEST sample of 66 galaxies containing 19,999 kpc-scale star-forming spaxels.  The main distinguishing feature of the extended ALMaQUEST sample is its diversity, since it contains galaxies on, below and above the main sequence, as well as central starbursts and mergers.  This diverse sample allows us to investigate how the star formation scaling relations respond to a broad range of physical conditions.  For the first time, we include the resolved dynamical equilibrium pressure (rPDE) relation amongst those investigated for the ALMaQUEST dataset. We compare the rPDE relation in the extended ALMaQUEST sample to the prediction from the pressure-regulated feedback-modulated (PRFM) theory of star formation (e.g. Ostriker \& Kim 2022).

\medskip

Our main conclusions are as follows.

\begin{itemize}

  \item  \textbf{Star formation scaling relations in the extended ALMaQUEST sample: } In addition to the strong scaling relations seen in the full sample of star-forming spaxels (as found in previous papers in the ALMaQUEST series), the rSK relation, the rSFMS and the rMGMS all persist even in galaxies experiencing more extreme conditions, such as in mergers and those with central starbursts (Figures \ref{ks_fig} -- \ref{sfms_fig}).

\medskip
    
\item  \textbf{The resolved $P_{DE}$ relation in the extended ALMaQUEST sample and comparison with the pressure regulated feedback modulated model: }  There is a strong correlation between $P_{DE}$ and \sigsfr, as expected from the PRFM model.  However, rather than a linear relation, the data exhibit a flattening at high values of $P_{DE}$ that is not predicted by the model (Figure \ref{prfm_fig}).  Although previous datasets only sparsely sample the high pressure regime, both the PHANGS and EDGE-CALIFA surveys show similar deviations from the PRFM model (Figure \ref{lit_fig}).

\medskip

\item  \textbf{Diversity in the rPDE relation: } There is significant galaxy-to-galaxy variation in the rPDE relation (Figure \ref{prfm_pifu}).  By dissecting the extended ALMaQUEST sample into different subsets we find that the rPDE relation behaves differently depending on the global SFR of the galaxy.  Galaxies that lie within $\pm$0.3 dex of the global SFMS show good agreement with the PRFM model (as has been previously found for other samples of normal star-forming galaxies in the nearby universe), middle panel of Figure \ref{pde_dsfr}.    Conversely, galaxies above and below the main sequence show significant offsets from the PRFM (top and bottom panels of Figure \ref{pde_dsfr}).   Moreover, we find that the magnitude and direction of the offset between the rPDE of a given galaxy and the PRFM prediction is directly anti-correlated with $\Delta$SFR (Figure \ref{dpde_dsfr}).   The flattening of the rPDE relation in our full sample (top left panel of Figure \ref{prfm_fig}) can therefore be understood by the superposition of normal star-forming galaxies and high SFR galaxies, that each contribute $\sim$ 50\% of the sample (top and middle panels of Figure \ref{pde_dsfr}).

    \medskip
    
\item \textbf{Uncertainties in the data:} We discuss a range of possible caveats in our calculation of $P_{DE}$.  Although the precise value of $\alpha_{CO}$ is likely to be the largest source of uncertainty, adopting various alternative prescriptions (e.g. metallicity or density dependent conversion factors) does not reconcile our data with the predictions from the PRFM model.
    
\medskip

    \item  \textbf{Dynamical equilibrium pressure as a star formation rate regulator:  }  We compare the scatters in the four star formation scaling relations and find that they vary depending on the sample of galaxies chosen, but are typically 0.2 -- 0.3 dex; the rPDE relation is no tighter than any of the other relations (Figure \ref{rms_fig}).  However, given the clear non-linearity of the observed rPDE relation, a comparison to a linear fit is inadequate to properly quantify the relationship between $P_{DE}$ and \sigsfr.  A random forest analysis, which is capable of capturing non-linear dependences in the data, shows unambiguously that  $P_{DE}$ is a better predictor of \sigsfr\ than either \sigh2\ or \sigstar\ alone (Figure \ref{rf_fig}).  In this sense, $P_{DE}$ is more `fundamental' for regulating star formation than \sigstar\ or \sigh2\ alone (even though the form of this relation deviates from the PRFM theory).

  \end{itemize}

The work presented here thus extends previous comparisons of the PRFM model in normal star-forming disks (where it has been seen to work well, e.g. Herrera-Camus et al. 2017; Sun et al. 2020a, 2023; Barrera-Ballesteros et al. 2021a) into more extreme environments, where it apparently breaks down.  Conversely, the data are qualitatively consistent with a model in which inflows of gas contribute to (and potentially dominate) the ISM turbulence (Krumholz et al. 2018) when SFRs are elevated above fiducial SFMS values (Figure \ref{k18_fig}).  

\section*{Acknowledgements}

The authors acknowledge an NSERC Discovery Grant (SLE), MOST 108-2628-M-001-001-MY3 and NSTC 111-2112-M-001-044- (LL), National Science and Technology Council of Taiwan grant 110-2112-M-032-020-MY3 (HAP), Australian Research Council Laureate Fellowship award FL220100020 (MRK), DGAPA-PAPIIT/UNAM grant IA-101522 and CONACYT grant CF19-39578 (JBB) and National Science Foundation Grant No 2205551 (JMS).

SLE is grateful to the European Southern Observatory, the University of Bologna and INAF Arcetri Observatory for funding extended visits during which work on this project took place.  The stimulating environments of these institutes was critical for the conception and development of this research.  In particular, SLE thanks Munan Gong for seeding the idea of comparing the ALMaQUEST to the PRFM model - it was a productive coffee time chat!  We are also grateful to Eve Ostriker and Chang-Goo Kim for discussions of the PRFM model and for sharing the formalism of their forthcoming work (Hassan et al. in prep) in advance of publication.  Conversations with Yu-Hsuan (Eltha) Teng and Adam Leroy also contributed significantly to the discussion presented in this work.

This paper makes use of the following ALMA data:  ADS/JAO.ALMA\#2015.1.01225.S, ADS/JAO.ALMA\#2017.1.01093.S, ADS/JAO.ALMA\#2018.1.00558.S, ADS/JAO.ALMA\#2018.1.00541.S, ADS/JAO.ALMA\#2019.1.00260.S.

ALMA is a partnership of ESO (representing its member states), NSF (USA) and NINS (Japan), together with NRC (Canada), MOST and ASIAA (Taiwan), and KASI (Republic of Korea), in cooperation with the Republic of Chile. The Joint ALMA Observatory is operated by ESO, AUI/NRAO and NAOJ.  The National Radio Astronomy Observatory is a facility of the National Science Foundation operated under cooperative agreement by Associated Universities, Inc.  Funding for the SDSS IV has been provided by the Alfred P. Sloan Foundation, the U.S. Department of Energy Office of Science, and the Participating Institutions. SDSS-IV acknowledges support and resources from the Center for High Performance Computing at the University of Utah.

\section*{Data Availability}

The data used in this paper are all released in a supplementary table in the electronic version of this paper.

\end{document}